\documentclass[a4paper,12pt]{article}
\pdfoutput=1
\usepackage{feynmp-auto,expdlist}
\usepackage{amsmath, amsfonts, amssymb}
\usepackage{graphicx}
\usepackage{enumerate}
\usepackage{latexsym}
\usepackage{hepnicenames}
\usepackage{enumerate}
\usepackage{soul}
\usepackage[normalem]{ulem}
\usepackage{bbold}
\usepackage{wasysym} 
\usepackage{makecell}
\RequirePackage[ colorlinks=true,urlcolor=blue,anchorcolor=blue,citecolor=blue,filecolor=blue,
               linkcolor=blue,menucolor=blue,linktocpage=true,pdfproducer=medialab,pagebackref=true]{hyperref}
 \hypersetup{  
 colorlinks   = true, 
 urlcolor     = ForestGreen, 
 linkcolor    = blue, 
  citecolor   = blue 
}

\newcommand{\blue}[1]{\color{blue}#1\color{black}}

\font\tenrsfs=rsfs10 at 12pt
\font\sevenrsfs=rsfs7
\font\fiversfs=rsfs5
\newfam\rsfsfam
\textfont\rsfsfam=\tenrsfs
\scriptfont\rsfsfam=\sevenrsfs
\scriptscriptfont\rsfsfam=\fiversfs

\numberwithin{equation}{section}

\usepackage{mathrsfs}
\usepackage{braket}
\usepackage{titling}
\usepackage{amsmath}
\usepackage{slashed}
\usepackage{amssymb}
\usepackage{epsfig}
\usepackage{graphicx}
\usepackage{color}
\usepackage{rotating}
\usepackage[margin=1.in]{geometry}
\usepackage[table,xcdraw,dvipsnames]{xcolor}
\usepackage[compress,numbers,sort]{natbib}
\usepackage{colortbl}
\usepackage{pdflscape}
\usepackage{color}
\usepackage{mathtools}
\usepackage{colortbl}
\definecolor{Gray}{gray}{0.95}
\definecolor{RGray}{gray}{0.85}
\definecolor{CGray}{gray}{0.92}

\newcommand{\N}{{\cal N}}

\definecolor{nicered}{rgb}{0.7,0.1,0.1}
\definecolor{nicegreen}{rgb}{0.1,0.5,0.1}
\definecolor{red}{rgb}{1.0, 0, 0}
\definecolor{niceblue}{rgb}{0,0,0.8}
\definecolor{red}{rgb}{1.0, 0, 0}
\allowdisplaybreaks

\definecolor{rosso}{cmyk}{0,1,1,0.4}
\definecolor{rossos}{cmyk}{0,1,1,0.55}
\definecolor{rossoc}{cmyk}{0,1,1,0.2}
\definecolor{blu}{cmyk}{1,1,0,0.3}
\definecolor{blus}{cmyk}{1,1,0,0.6}
\definecolor{bluc}{cmyk}{1,1,0,0.1}
\definecolor{verde}{cmyk}{0.92,0,0.59,0.25}
\definecolor{verdec}{cmyk}{0.92,0,0.59,0.15}
\definecolor{verdes}{cmyk}{0.92,0,0.59,0.4}

\def\eq#1{{Eq.~(\ref{#1})}}
\def\eqs#1#2{{Eqs.~(\ref{#1})--(\ref{#2})}}
\def\fig#1{{Fig.~\ref{#1}}}

\def\sect#1{{Sect.~\ref{#1}}}

\renewcommand{\bar}{\overline}

\newcommand{\beq}{\begin{equation}}
\newcommand{\eeq}{\end{equation}}
\newcommand{\bea}{\begin{eqnarray}}
\newcommand{\eea}{\end{eqnarray}}

\renewcommand{\[}{\left[}
\renewcommand{\]}{\right]}
\renewcommand{\(}{\left(}
\renewcommand{\)}{\right)}

\newcommand{\LQCD}{T_{\text{QCD}}}
\usepackage[capitalise]{cleveref}
\usepackage{nicefrac}
\usepackage{subcaption}
\newcommand{\email}[1]{\href{mailto:#1}{\tt #1}}

\begin{document}
\vspace{1.5cm}

{\flushright
{\blue{ \hfill}\\
\blue{ \hfill}\\
\blue{DESY 21-011}\\
\blue{IFT-UAM/CSIC-20-144}\\
\blue{FTUAM-20-21}}\\
\hfill 
}

\begin{center}
{\LARGE\bf\color{blus} 
 Dark matter from an even lighter QCD axion: trapped misalignment}\\[1cm]

{\bf Luca Di Luzio$^{a}$, Belen Gavela$^{b,\,c}$, Pablo Quilez$^{a}$, Andreas Ringwald$^{a}$}\\[7mm]

{\it $^a$Deutsches Elektronen-Synchrotron DESY, \\ 
Notkestra\ss e 85, D-22607 Hamburg, Germany}\\[1mm]
{\it $^b$Departamento de Fisica Teorica, Universidad Autonoma de Madrid, \\ 
Cantoblanco, 28049, Madrid, Spain}\\[1mm]
{\it $^c$Instituto de Fisica Teorica, IFT-UAM/CSIC, \\
Cantoblanco, 28049, Madrid, Spain}\\[1mm]

\vspace{0.5cm}

\begin{abstract}
We show that dark matter can be accounted for by an axion that solves the strong CP problem, but is much lighter than usual due to a $Z_\N$ symmetry. 
The whole mass range  from the canonical QCD axion down to the ultra-light regime  is allowed, with $3\le\N\lesssim65$. This includes the first proposal of a ``fuzzy dark matter'' QCD axion with $m_a\sim10^{-22}$ eV.
A novel misalignment mechanism occurs --{\it trapped misalignment}-- due to the peculiar temperature dependence of the $Z_{\N}$ axion potential. The dark matter  relic density is enhanced because the axion field undergoes two stages of oscillations: it is first trapped in the wrong minimum, which effectively delays the onset of true oscillations. Trapped misalignment is more general than the setup discussed here, and may hold whenever an extra source of Peccei-Quinn breaking appears at high temperatures. Furthermore, 
  it will be shown that trapped misalignment  can 
   dynamically   source the recently proposed kinetic misalignment mechanism. 
  All the parameter space is within tantalizing reach of the experimental projects for the next decades. For instance, even Phase I of 
CASPEr-Electric could discover this axion.

\end{abstract}

\begin{minipage}[l]{.9\textwidth}
{\footnotesize \vspace{1cm}
\begin{center}
\textit{E-mail:} 
\email{luca.diluzio@desy.de}\,,
\email{belen.gavela@uam.es}\,, \\
\email{pablo.quilez@desy.de}\,,
\email{andreas.ringwald@desy.de}
\end{center}}
\end{minipage} 
\thispagestyle{empty}
\bigskip

\end{center}

\setcounter{footnote}{0}

\newpage
\tableofcontents

\newpage 

\section{Introduction } 
\label{sec:intro}

Were an  axion-like particle (ALP)  to be ever discovered, it would be 
 compelling to explore whether it has something
to do with the strong CP problem~\cite{Peccei:1977hh,Peccei1977,Weinberg:1977ma,Wilczek1978}, and/or whether it can be a viable dark matter 
(DM) candidate~\cite{Preskill:1982cy,Abbott:1982af,Dine:1982ah}. 
The canonical QCD axion (aka ``invisible axion'')  satisfies 
  \begin{equation}
  m_a^{\rm QCD} f_a = m_\pi \, f_\pi\,
  \frac{\sqrt{z}}{(1+z)}\,, \qquad \text{where}\quad z=\frac{m_u}{m_d}\, 
  \label{canonical}
  \end{equation}
   and  $m_a^{\rm QCD}$,  $f_a$, $f_\pi, m_\pi,  m_u$ and $m_d$  denote respectively  the axion mass, the axion scale, the pion decay constant, and the pion, up  and down quark masses~\cite{Kim:1979if,Shifman:1979if,Zhitnitsky:1980tq,Dine:1981rt,Kim:1984pt,Choi:1985cb}. Eq.~(\ref{canonical}) holds whenever QCD is the only confining group of the theory, 
   irrespectively of the ultraviolet (UV)  model details.  A departure from this $m_a$-$f_a$ relation  always requires to extend the gauge confining sector beyond the Standard Model (SM) QCD group.
  
Axions that solve the strong CP problem but are heavier than the canonical QCD axion have been explored since long and  revived in the last years~\cite{rubakov:1997vp,Berezhiani:2000gh,Gianfagna:2004je,Hsu:2004mf,Hook:2014cda,Fukuda:2015ana,Chiang:2016eav,Dimopoulos:2016lvn,Gherghetta:2016fhp,Kobakhidze:2016rwh,Agrawal:2017ksf,Agrawal:2017evu,Gaillard:2018xgk,Buen-Abad:2019uoc,Hook:2019qoh,Csaki:2019vte,Gherghetta:2020ofz}.   In contrast, solutions to the strong CP problem with lighter axions were uncharted territory until very recently.

The goal of this work is to determine the implications for DM of a freshly proposed dynamical --and technically natural-- scenario~\cite{Hook:2018jle,ZNCPpaper}, which solves the strong CP problem with an axion much lighter than the canonical QCD one.  
Its key point was  to assume that Nature is endowed with a $Z_\N$ symmetry  realized non-linearly by the axion field~\cite{Hook:2018jle,ZNCPpaper}. $\N$ mirror and degenerate worlds  linked by the axion field would coexist with the same coupling strengths as in the SM, with the exception of the 
effective $\theta_k$-parameters, 
\begin{equation}
\label{Eq: Lagrangian ZN}
\mathcal{L} 
= 
\sum_{k=0}^{\N-1} \[ \mathcal{L}_{\text{SM}_k} +
\frac{\alpha_s}{8\pi}  \( \theta_a + \frac{2 \pi k}{\N} \) G_k \widetilde G_k \] \,+ \dots \,.
\end{equation}
Here, $\mathcal{L}_{\text{SM}_k}$ denotes  exact  copies of the SM total Lagrangian excluding the  
topological $G_k\tilde G_k$ term,\footnote{The SM is identified from now on with the $k=0$ sector and this label on SM quantities 
will be often dropped in the following. } 
$\theta_a \in [-\pi,\pi)$ is defined in terms of the axion field $a$,  
$\theta_a\equiv a/f_a$,  and  the dots stand for  $Z_\N$-symmetric 
portal 
couplings among the SM copies.

 The solution to the strong CP problem of this $Z_\N$ scenario required  $\N$ to be odd. 
 Overall, the $\sim 10$ orders of magnitude tuning required by the SM strong CP problem is traded by a $1/\N$ adjustment, where $\N$ could be as low as $\N=3$ (viable solutions were found with $3\le\N\lesssim47$ for $f_a\lesssim 2.4\times 10^{15}$ GeV, and any $\N$ for larger 
 values of $f_a$~\cite{ZNCPpaper}). 
 The resulting axion is exponentially lighter than the canonical QCD one, because the non-perturbative contributions to its potential from the $\N$ degenerate QCD groups  conspire by symmetry to suppress each other.  
  This can be intuitively understood from the large $\N$ limit of a non-linearly realized 
  $Z_\N$ shift symmetry, which is a continuous global $U(1)$ symmetry: the 
  axion acts as the Goldstone boson of the discrete symmetry~\cite{Das:2020arz}.   Its mass thus  vanishes asymptotically for large $\N$  as befits a $U(1)$ 
   Goldstone boson. Indeed, it has been shown~\cite{ZNCPpaper} that in the large $\N$ limit the total axion potential is given in all generality by
  \beq 
  \label{Eq: fourier potential large N hyper}
V_{\mathcal{N}}\left(\theta_a\right)
  \simeq - \frac{m_a^2 f_a^2}{\N^2} \,\cos (\N\theta_a)\,, 
\eeq 
where now the axion mass in vacuum obeys
\beq
m_a^2 \simeq \frac{m_{\pi}^2 f_{\pi}^2}{f_a^2} \frac{1}{\sqrt{\pi}} \sqrt{\frac{1-z}{1+z}} \,\,\mathcal{N}^{3 / 2} \, z^{\mathcal{N}} \,,
\label{maZNLargeN}
\eeq
 which  is  exponentially suppressed ($\propto z^\N$)  in comparison to  $(m_a^{\rm QCD})^2$.\footnote{Note that the value of $m_a$ for $\N=1$ should be equal to $m_a^{\rm QCD}$ given in Eq.~(\ref{canonical}), while Eq.~(\ref{maZNLargeN}) only holds in the large $\N$ limit.} 
 
     The crucial properties of such a light axion are generic and do not depend on the details of the putative UV completion. An important byproduct of this construction  is  an enhancement of all axion interactions which is {\it universal}, that is, model-independent and equal for all axion couplings, at fixed $m_a$.  The detailed exploration  of the $Z_\N$ paradigm and of the phenomenological constraints which do not require the axion to account for DM can be found in Ref.~\cite{ZNCPpaper}.
  
   We will determine here instead the  parameter space that solves both the strong CP problem  and the nature of DM within the $Z_\N$ framework under discussion.   The question of DM  is   of strong  experimental interest and very timely,  
 in particular given the plethora of projects  targeting light DM candidates, down to the region of ``fuzzy'' DM~\cite{Hui:2016ltb}. For instance,  
CASPEr-Electric~\cite{Budker:2013hfa,JacksonKimball2017,JacksonKimball2020a} probes directly the anomalous gluonic  coupling of the axion via  an oscillating  neutron electric dipole moment (nEDM): the strength of that coupling cannot be modified for the canonical QCD axion irrespective of the model details, unlike  other axion-to-SM couplings that can be selectively enhanced~\cite{DiLuzio:2016sbl,Farina:2016tgd,DiLuzio:2017pfr,Agrawal:2017cmd,Marques-Tavares:2018cwm,DiLuzio:2020wdo,Darme:2020gyx}. 
  We show here that, in contrast,   a hypothetical early discovery at  
   CASPEr-Electric could be interpreted as a solution to the strong CP problem via a $Z_\N$ reduced-mass axion, because of the same-size enhancement of {\it all} axion couplings 
   in such a scenario. Note that an exceptionally light QCD axion has formerly been considered~\cite{Alvarez:2017kar}, albeit it required an ad-hoc  fine-tunning  of  potential parameters. 
   
It will be shown that the cosmology of the $Z_\N$ axion exhibits several novel aspects.  
The cosmological impact of hypothetical parallel ``mirror''  worlds has been studied at length in the literature (for a review, see e.g.~\cite{Berezhiani:2003xm}). 
In particular, the constraints on the number of effective relativistic species $N_{\text{eff}}$ present in the early Universe imply that the mirror copies of the SM must be less populated --cooler-- than the ordinary SM world \cite{Berezhiani:2000gw}. 
This requires in turn that the SM has never been in thermal contact with its mirror replica. Fortunately, mechanisms that source this world-asymmetric initial temperatures while preserving the 
$Z_\N$ symmetry may arise naturally in the cosmological 
evolution~\cite{Kolb:1985bf,Berezinsky:1999az,Dvali:2019ewm}.  
It has also been suggested that DM could be simply constituted by mirror matter, 
for which  relevant constraints apply given the differences in temperatures. 
Note, nevertheless, that in most cases only one replica of the SM was considered, 
while large $\N$ could significantly modify those analyses.  
Furthermore, an axionic nature of DM has not been previously considered in the mirror world setup with a reduced-mass axion.\footnote{An axion {\it heavier} than the QCD one has been contemplated  as DM, within a $Z_2$ setup which realized linearly the symmetry~\cite{Giannotti:2005eb}.} 
While gravity and axion-mediated interactions are naturally small enough, the impact of other possible ($Z_\N$-symmetric)  interactions on the thermal communication between the SM and its mirror copies  will be discussed.

   The evolution of the $Z_\N$ axion field 
and its contribution to the DM relic abundance 
 will be shown to 
depart drastically from both the standard case and the previously considered mirror world scenarios. 
   Due to the peculiar temperature dependence of the $Z_{\N}$ axion potential, the production of DM axions in terms of the misalignment mechanism is modified. The scenario results in a novel type of misalignment, with a large value of the misalignment angle. 
 In particular, we will show that the relic density is enhanced because the axion field undergoes two stages of oscillations, 
  that are separated by 
  an unavoidable and 
  drastic --non-adiabatic-- modification of the potential. The axion field
  is first \emph{trapped} in the wrong minimum (with $\theta=\pi$), which effectively delays the onset of the true oscillations and thus enhances the DM density.\footnote{While this manuscript was being written, Ref.~\cite{Nakagawa:2020zjr} appeared  
  where a trapping mechanism was used
  in a different context, i.e.~to {\it reduce} the axion DM abundance. Conversely, an alternative enhancement mechanism has also been  proposed recently \cite{Alonso:2020com}.}   
 We will call this new production mechanism \emph{trapped misalingment}. While an early stage of oscillations has been previously proposed in the literature in order to adiabatically suppress the axionic energy density and/or to solve the isocurvature problem \cite{Linde:1996cx,Fischler:1983sc,Nomura:2015xil,Kawasaki:2017xwt}, our trapped misalingment differs from previous works in that the modification of the potential (from the high temperature large axion mass to the exponentially suppressed zero temperature mass) is strongly non-adiabatic and, when active, always leads to an enhancement of the relic density.  
Note that inflation played a crucial role in previous dynamical setups advocated  to drive the initial misalignment to 
$\pi$~\cite{Co:2018mho,Takahashi:2019pqf,Huang:2020etx}, 
while our mechanism  results directly from temperature effects. 

Furthermore it will be shown that, in some regions of the parameter space,  trapped misalignment will automatically source the recently proposed kinetic misalignment mechanism~\cite{Co:2019jts}.  
In the  latter, a sizeable initial axion velocity is the source of the axion relic abundance as opposed to the conventionally assumed initial misalignment angle. The original kinetic misalignment proposal~\cite{Co:2019jts} required an ad-hoc Peccei Quinn (PQ)-breaking non-renormalizable effective 
operator suppressed by powers of the Planck mass, $M_{\rm Pl}$.  In contrast, we will show that the early stage of oscillations in the $Z_\N$ axion framework naturally flows out into  kinetic misalignment.

The interplay of the different mechanisms is studied in detail. Moreover, although general phenomenological consequences of a large misalignment angle were studied in  Ref.~\cite{Arvanitaki:2019rax}, we will identify the main novel consequences that follow from the scenario under discussion.  
 On the phenomenological arena, we discuss the implications of the $Z_\N$ reduced-mass axion  
for axion DM searches, namely those experiments that rely on the 
hypothesis that an axion or ALP sizeably contributes to the DM density.  More specifically, the experimental 
prospects to probe its coupling to photons, 
nucleons, electrons and the nEDM operator are considered.  The analysis sweeps through the whole mass range down to ultra-light axions (with masses $m_a \ll 10^{-10}$ eV), within the $Z_\N$ axion framework under discussion.  The present and projected sensitivity to the number of possible mirror world $\N$ will be determined, and the constraints obtained will be compared and combined with those stemming from experiments which are independent of whether axions account or not for DM~\cite{ZNCPpaper}.

The structure of the  paper can be easily inferred from the Table of Contents.

\section{Cosmological constraints on mirror worlds}
\label{sec:cosmo constraints}
The existence of a hypothetical parallel mirror world, with microphysics identical to that of the observable world and connected to the latter only gravitationally, has a non-trivial impact on cosmology (for a review, see e.g.~\cite{Berezhiani:2003xm}). 
In particular, extra relativistic species in the early Universe (mirror photons, electrons and neutrinos) affect the number of effective neutrino species  
$N_{\rm eff}$  that can be measured through: i) the abundances of light elements  --in particular Helium--  at   
the time of Big Bang nucleosynthesis (BBN), i.e.~at $T\sim 1\, {\rm MeV}$; ii) through the damping tail in the cosmic microwave background (CMB) power spectrum at recombination,  $T\sim 0.26\, {\rm eV}$. 
  Present data yield  \cite{Fields:2019pfx,Akrami:2018vks,Aghanim:2018eyx}\footnote{For BBN, the constraint corresponds to $BBN+Yp+D$ in Table V of Ref.~\cite{Fields:2019pfx}, translating their 68.3\% CL result for the number of neutrino species $N_\nu$
 to a 95\% CL value $N_{\nu}=2.85 \pm 0.28$, and $N_{\mathrm{eff}}=1.015 \, N_{\nu}$. 
For CMB, the 
  combination of TT+TE+EE+lowE+lensing+BAO data by Planck 2018~\cite{Akrami:2018vks,Aghanim:2018eyx} is considered here; note that due to the disagreement between local and CMB measurements of the Hubble expansion parameter $H_0$, the constraint on $\Delta N_{\rm eff}$ would weaken  to $N_{\mathrm{eff}}=3.27 \pm 0.30$ if  $H_0$ measurements were included.
} 
\begin{equation}
  \text{BBN: } N_{\text{eff}}=2.89 \pm 0.57 \,\,,\qquad 
   \text{CMB: } N_{\text {eff }}=2.99_{-0.33}^{+0.34}\,.
  \label{Eq: BBN neff bound}
\end{equation} 
In order to satisfy these constraints in the presence of one mirror world,  its temperature $T'$   needs to be smaller than that of the SM, $T_{0}$. 
 In the $Z_\N$ scenario under discussion the different $k > 0$ copies of the SM could have different temperatures $T_{k} <T_{0}$,  and hence different energy densities $\rho_k$ and entropies $s_k$,
\begin{align}
\label{eq:rhoksk}
\rho_k=\frac{\pi^2}{30}{g}_* (T_k) T_k^4\,,\qquad s_k=\frac{2\pi^2}{45}{g}_{s}(T_k) T_k^3\,,
\end{align}
where $g_*$ and $g_s$ denote the effective degrees of freedom related to energy  and entropy density, respectively. 
 In principle, the functional forms 
$g_*, g_s$ could have an extra 
$k$-dependence through the respective  $\theta_k$ parameters, but the 
 overall impact is expected to be minor and  will be disregarded  in the following. 
Moreover, we  assume  here that thermally produced axions give 
a negligible contribution to $N_{\rm eff}$. This is typically 
the case for $f_a \gtrsim 10^{7\div 8}$ GeV \cite{Chang:1993gm,Hannestad:2005df,DiLuzio:2021vjd} 
and if model-dependent axion couplings to SM fermions are neglected \cite{DEramo:2018vss}.

The different sectors evolve in time with separately conserved entropies, therefore the ratio of entropy densities $\gamma^3_k\equiv(s_k/s)$ is constant, while the ratio of temperatures is given by
\begin{align}
\frac{T_k}{T}=\gamma_k \cdot\bigg[\frac{g_{s}(T)}{g_{s}\left(T_k\right)}\bigg]^{1 / 3}\equiv \gamma_k \cdot b_k(T,T_k) ,\qquad \text{for }\, k\equiv1,\dots ,\N-1\,, 
\label{gammalong}
\end{align}
  where from now on $T$ will denote the  temperature of the SM copy, $T \equiv T_0$. 
The Hubble expansion rate depends on the total energy density of the Universe. In a radiation dominated era with $\N$ mirror worlds it reads 
\beq
H^2(t)=\frac{4\pi^3}{45} \sum_{k=0}^{\N-1} {{g}_{*}(T_k)} \frac{T_k^{4}}{M^2_{\rm Pl}} 
\equiv \frac{4\pi^3}{45} \frac{T^{4}}{M^2_{\rm Pl}}\,{\bar{g}_{*}(T)} \, ,
\label{Eq:friedmann equation} 
\eeq
with the number of effective degrees of freedom $\bar{g}_{*}(T)$  given by 
\begin{equation}
  \bar{g}_{*}(T)=g_{*}(T)\left(1+\sum_{k=1}^{\N-1} c_k\gamma_k^{4}\right)\,,
  \label{Eq: Zn number of rel dof g*}
\end{equation}
and  $c_k\left(T, T_k\right)\equiv \left[g_{*}\left(T_{k}\right) / g_{*}(T)\right] \cdot\left[g_{s}(T) / g_{s}\left(T_{k}\right)\right]^{4 / 3}$. 

Let us discuss next the implications for $N_{\rm eff}$.  In the SM,   $g_*(T=1\, {\rm MeV})=5.5+\frac{7}{4}N^{\rm SM}_{\rm eff}=10.83$ where $N^{\rm SM}_{\rm eff}=3.046$  is the effective number of neutrino species. 
Analogously, at recombination  $g_*(T=0.26\,{\rm eV})=3.93 $.
 It follows from \cref{Eq: Zn number of rel dof g*} that the contribution of the ensemble of $\N$ worlds to $\Delta N_{\rm eff} = N_{\rm eff} - N^{\rm SM}_{\rm eff}$ at BBN and CMB is given by 
\begin{align}
  \text{BBN: }\quad \Delta N_{\rm eff} =\frac{4}{7}g_*(T=1\, {\rm MeV}) \sum_{k=1}^{\N-1}c_k \gamma_k^{4} 
  \, \quad\Longrightarrow \quad\sum_{k=1}^{\N-1}c_k \gamma_k^{4}<0.067
\label{Eq: Neff BBNMeV} 
  \, ,\\
  \text{CMB: }\quad \Delta N_{\rm eff} =\frac{4}{7}g_*(T=0.26\,{\rm eV}) \sum_{k=1}^{\N-1}c_k \gamma_k^{4} 
  \, \quad \Longrightarrow \quad \sum_{k=1}^{\N-1}c_k \gamma_k^{4}<0.13\,,
\label{Eq: Neff BBNeV}
\end{align}
where the bounds on the right-hand side stem from  the  constraints in 
Eq.~(\ref{Eq: BBN neff bound}).
 These expressions illustrate an interesting prediction of the mirror world(s) scenario: {\it  the deviation in the number of measured effective neutrino species  with respect $N^{\rm SM}_{\rm eff}$ is not constant but varies with  temperature.}  
 That is, it may vary with time as certain mirror species will become non-relativistic at different times in the evolution, 
 because of the different temperatures of the mirror worlds. 

\subsubsection*{$c_{k}\simeq 1$ approximation} 
The temperature dependence of  the function ${g}_{s}$ is quite flat in most regions of interest, ${g}_{s}(T_{k\ne0})\simeq{g}_{s}( T)$. Within this approximation $b_{k\ne0}\simeq c_{k\ne0}\simeq 1$, and the parameter $\gamma_k$ directly 
gives the ratios of temperatures  
\begin{equation}
\gamma_k \simeq \frac{T_k}{T}\,,
\label{ak1approx}
\end{equation}
which remain constant throughout the Universe evolution, with $\gamma_k<1$ for all $k\ne 0$.  This approximation is valid as far as the temperatures involved correspond to the same plateau of the $N_{\rm eff}$ distribution (see e.g.~\cite{Baumann:2018muz}). In other words, as long as the number of relativistic degrees of freedom does not change between $T_k$ and $T$ and thus $c_{k}\simeq 1$ holds. This is the case at the CMB temperature, where Eq.~(\ref{ak1approx}) holds for all $k$ because all exotic worlds are cooler than the SM one. 
For temperatures around the BBN region, we have checked that the approximation is still good as long as 
$\gamma_k$ is not too small, e.g.~better than $25\%$ for $\gamma_{k\ne0}\gtrsim 0.1$, which does not change noticeably the analytical expressions. We will work in this framework all through the rest of the paper, except for the numerical results.  More details can be found in Appendix~\ref{App: ratios gs grho}. 

 In general, the $\N-1$ copies of the SM may have different temperatures. In this  case  Eqs.~(\ref{Eq: Neff BBNMeV})-(\ref{Eq: Neff BBNeV})
 allow to constrain the temperature of the hottest of them,  ${T_{\rm max}}$, 
 \begin{equation}
  \text{BBN:} \quad \frac{T_{\rm max}}{T}<0.51\,, \qquad \text{CMB:}  \quad \frac{T_{\rm max}}{T}<0.60\,.
  \label{Eq: BBN neff bound Hot}
\end{equation} 
  If all mirror images of the SM are  instead assumed to have the same temperature,   $T_{k}\sim T'$ for $k\ne 0$, 
  the most restrictive bounds  follow,  
 \begin{equation}
  \text{BBN:}  \quad \frac{T'}{T}<\frac{0.51}{(\N-1)^{1/4}}\,, \qquad \text{CMB:}  \quad \frac{T'}{T}<\frac{0.60}{(\N-1)^{1/4}}\,.
  \label{Eq: BBN neff bound3}
\end{equation}
 It is interesting that BBN data set the most constraining bound on  $N_{\rm eff}$,  in spite of CMB measurements being more precise. This is due 
   to the temperature dependence of the mirror worlds contribution. Moreover, this non-trivial dependence represents a smoking gun for the existence of the latter, as it  generically predicts incompatible measurements of $N_{\rm eff}$ from BBN and CMB. Specifically, the scenario predicts in all generality the following discrepancy:
 \begin{align}
 N^{\rm BBN}_{\rm eff}- N^{\rm CMB}_{\rm eff}=3.92 \sum_{k=0}^{\N-1}c_k \gamma_k^{4}\,\simeq\,3.92 \sum_{k=0}^{\N-1} \gamma_k^{4}\,.
 \end{align}
 If such a difference were to be experimentally established, it would allow  to predict the temperature of the mirror worlds e.g.~in the two limiting cases in \eqs{Eq: BBN neff bound Hot}{Eq: BBN neff bound3}.

\subsubsection*{Constraints on portal couplings} 
The SM must avoid thermal contact with its mirror copies all through the (post-inflation) history of the Universe so as to fullfil 
the condition $T_{k\neq 0} \ll T$. 
 This implies that the interactions between the SM and its  copies 
need to be very suppressed. Non-renormalizable interactions such as 
gravity and axion-mediated ones are naturally small enough, while the 
Higgs and  hypercharge kinetic portal couplings can potentially spoil the 
condition $T_{k\neq 0} \ll T$. 
For instance, in the $Z_2$ mirror case $T' / T \lesssim 0.5$ requires 
both portal couplings, defined as $\mathcal{L} \supset \kappa |H|^2 |H'|^2 + \epsilon B^{\mu\nu} B'_{\mu\nu}$,   where $H$ and $H'$ ($B^{\mu\nu}$ and $B'_{\mu\nu}$) denote respectively the SM Higgs doublet (hypercharge field strength) and its mirror copy, and $\kappa$  and $\epsilon$ are dimensionless couplings, to respect  $\kappa \,, \epsilon \lesssim 10^{-8}$~\cite{Berezhiani:2005ek}.   
Even smaller couplings are needed  in the  $Z_\N$ case with $\N>2$. This can suggest a `naturalness' issue for the Higgs and 
  kinetic portal couplings, as they cannot be forbidden in terms of internal symmetries. Nevertheless, such small couplings  may be technically natural because of an enhanced \emph{Poincare} symmetry \cite{Volkas:1988cm,Foot:2013hna}: in the limit where non-renormalizable interactions are neglected,  the 
  $\kappa, \epsilon \to 0$ limit for the ensemble of portals corresponds to an enhanced $\mathcal{P}^\N$  
symmetry (namely, an independent space-time Poincar\'e transformation 
$\mathcal{P}$ in each sector).  This protects those couplings  from  radiative corrections other than those induced by 
the explicit $\mathcal{P}^\N$ breaking due to gravitational and axion-mediated interactions. 
The former are presumably small being Planck suppressed, while axion-mediated corrections 
to $\kappa$ scale like $m_H^2 / f^2_a$ \cite{Galda:2021hbr} and hence they can be safely neglected for the standard high $f_a$ values considered.
Moreover, 
axion-mediated interactions among the different sectors (leading to interaction rates scaling as $\sim T^5/f_a^4$) are also small enough during the evolution of the Universe, 
such that they do not spoil the evolution of the independent thermal baths, 
as long as the PQ breaking pre-inflationary scenario is considered.

\subsection{On asymmetric SM/mirror temperatures}
The microphysics responsible for the evolution of the 
early SM Universe  and of its mirror copies is almost the same.
 Which mechanisms can then source different temperatures for the SM and its replicae?

One difference in the microphysics of our setup is the axion coupling to the $G_{\mu\nu} \tilde G^{\mu\nu}$ pseudo-scalar density in \eq{Eq: Lagrangian ZN}: the effective value of the $\theta$ parameter differs for each sector $k$, $\theta_k = 2\pi k / \N$ (and thus relaxing to zero in the SM with probability $1/\N$ -- see \sect{sec:axion_dark_matter}). This  implies   that nuclear physics would be drastically different for the SM and its mirror copies. Indeed,  the one-pion scalar exchange parametrized by the effective baryon chiral Lagrangian,   
\beq 
\label{eq:LchiPTNucl}
\mathcal{L}_{\chi\text{PT}} 
\supset c_+  \sum_{k=0}^{\N-1} \frac{m_u m_d \sin\theta_k}{[m_u^2 + m_d^2 + 2 m_u m_d \cos\theta_k]^{1/2}} 
\frac{\pi_k^a}{f_\pi} \bar N_k \tau^a N_k \, ,
\eeq
where $\pi_k$ and $N_k$ denote respectively the pion and nucleon fields and $c_+$ is an $\mathcal{O}(1)$ low-energy constant, 
induces long-range 
forces (i.e.~not spin-suppressed in the non-relativistic limit)
among nucleons in all worlds but the SM one~\cite{Ubaldi:2008nf,Lee:2020tmi}. This may 
 lead to different thermalization histories of the mirror 
copies, modulated by their `distance' $k$ from the SM sector ($k=0$).
Qualitatively, nucleons are kept longer in thermal equilibrium in the mirror replicas 
of the SM due to this extra interaction channel, which could play a role in generating 
different temperatures. 
However, a quantitative estimate is complicated by the fact that 
the interaction in \eq{eq:LchiPTNucl} becomes relevant only around the QCD phase transition. 

Previously in the literature, mirror world scenarios leading to sufficiently different temperatures relied on specific inflation implementations. Some of them could also hold in our $Z_\N$ scenario. For instance,  in the so-called  \emph{asymmetric inflation} it is argued that  the initial condition after inflation may be very different for the SM and the various mirror copies, even if they have very similar microphysics dynamics.  
This mechanism was proposed in the context of an exactly $Z_2$ symmetric mirror world~\cite{Kolb:1985bf,Berezinsky:1999az}. 
Two inflaton fields would be present: $\phi$ for the SM and $\phi'$ for its $Z_2$ mirror copy, respectively slow-rolling towards 
the minimum of their potentials  (the generalization to the case of $Z_\N$ is straightforward). Inflation ends when the potential steepens 
and the inflaton field picks up kinetic energy so that it starts to oscillate towards the minimum 
of its potential. In this process, quantum fluctuations can be such that the two inflaton fields 
do not necessarily reach their minimum at the same time, even in the same spatial region. 
In regions where $\phi'$ reaches its minimum first, 
any mirror particles produced due to its oscillations are diluted by the remaining inflation driven 
by the second field $\phi$. Hence, by the time when $\phi$ reheats the SM sector, 
the density of mirror matter will  be very small, and in consequence $T' \ll T$. 
Quantitatively, whether such asymmetric inflation scenario can be realized or not 
depends on the shape of the inflaton potential~\cite{Berezinsky:1999az}. 
An alternative idea to achieve different temperatures is that the inflaton itself transforms non-linearly under the $Z_\N$ symmetry~\cite{Dvali:2019ewm}: this automatically implies different values for the inflaton fields in each mirror world, and thus different thermal histories. 
In the present paper we will not commit ourselves to a specific mechanism 
realizing different SM/mirror temperatures, 
but simply assume as a working hypothesis $T_{k \neq 0} < T$.

Mirror baryonic matter has been advocated to explain partially or even totally the nature of DM, in the context of the $Z_2$-symmetric mirror world (see e.g.~\cite{Berezhiani:2003xm}). This may be achieved, in spite of the lower temperature of the mirror world,  by means of a proper baryogenesis mechanism in the mirror sector \cite{Berezhiani:2000gw}. Nevertheless, the abundance of mirror baryons strongly depends on the specific baryogenesis mechanism at play and thus
in the following, we will not dwell with mirror baryonic DM (which is assumed to give a negligible contribution \footnote{For instance, the extension to our framework of one of the simplest baryogenesis scenarios studied in \cite{Kolb:1985bf} would generate negligible mirror baryon density as long as $\N \lesssim 3000$, which leaves unconstrained the parameter region considered in this paper.}
due to the lower temperature of the SM mirror copies) and focus instead on axionic DM.

\section{$Z_\N$ axion dark matter} 
\label{sec:axion_dark_matter}
\definecolor{orangeQCD}{rgb}{0.9725490196078431,0.6274509803921569,0.49411764705882355}
\definecolor{HubbleGreen}{rgb}{0.17254901960784313,0.6274509803921569,0.17254901960784313}
\definecolor{BlueTrapped}{rgb}{0.1568627450980392, 0.49019607843137253, 0.5568627450980392}
\definecolor{BlueKin}{rgb}{0.12941176470588237, 0.44313725490196076, 0.7098039215686275}
\definecolor{PurpleTrapped}{rgb}{0.7529411764705882, 0.21176470588235294, 0.3568627450980392}

\definecolor{colFinDen}{rgb}{0.894118,0.529412,0.317647}
\definecolor{colSNSup}{rgb}{0.137255,0.545098,0.270588}
\definecolor{colnEDM}{rgb}{0.031373,0.317647,0.611765}
\definecolor{colLab}{rgb}{0.066667,0.317647,0.172549}
\definecolor{colAstro}{rgb}{0.000000,0.427451,0.172549}
\definecolor{colAstroL}{rgb}{0.0, 0.66, 0.42}
\definecolor{colZN}{rgb}{0.968735,0.465621,0.212103}
\definecolor{purpleDMbands}{rgb}{0.623529,0.141176,0.384314}

One of the key ingredients in order to compute the axion relic density is the temperature at the onset of oscillations around the true minimum, since it determines for how long  axion oscillations are damped (and therefore to what extent  the relic density is diluted). In the traditional scenario, that corresponds to the temperature at which the Hubble parameter is of the order of the axion mass. 
We analyze here the intrinsic differences between the $Z_\N$ scenario with a reduced-mass axion and the canonical QCD axion.

An important question is whether the $Z_\N$ axion can account for the observed DM density, without requiring any further adjustment other than the inherent $1/\N$ tuning needed to solve the strong CP problem, i.e.~to choose $\theta_a=0$ as the vacuum for the SM world among the $\N$ possible  $\theta_a$ values,  $\theta_{a}=\{\pm 2 \pi \ell / \mathcal{N}\} \text { with } \ell=0,1, \ldots, (\mathcal{N}-1)/2$.
It will be demonstrated in this section that this is indeed the case.

We will focus here on the study of the axionic zero mode, assuming spatial homogeneity and the pre-inflationary PQ breaking scenario.
\subsection{Temperature dependence of the $Z_\N$ axion potential}
At temperatures well below the critical temperature of the QCD phase transition, 
$T \ll T_{\rm QCD} \simeq 150$ MeV, 
the customary QCD axion potential is well approximated by the zero temperature chiral potential.  However, the chiral expansion breaks down close to  $T_{\rm QCD} $.
  Lattice computations of the topological susceptibility $\chi (T)$  are instead available for the cross-over regime, from which the QCD axion mass is estimated.   As the exact temperature dependence is not known,   we will parametrize the QCD axion potential at finite temperature $V(\theta_a,T)$ 
   by weighing the zero temperature expression of the chiral axion potential $V(\theta_a)$ by a $h(T)$ factor
 \begin{equation}
 \label{QCDpotentialT}
  V(\theta_a,T)\equiv V(\theta_a)\,\times  h(T)=
 - m_{\pi}^{2} f_{\pi}^{2}\sqrt{1-\frac{4 m_{u} m_{d}}{\left(m_{u}+m_{d}\right)^{2}} \sin ^{2}\left(\frac{\theta_{a}}{2}\right)}\,\times  h(T)\,,
 \end{equation}
 where 
 \begin{equation}
\label{h}
  h(T)\equiv\left\{\begin{array}{ll}
 \,1 \qquad                         &   \text{ for } T<T_{\rm QCD} \\
\left(\frac{T_{\rm QCD}}{T}\right)^\alpha & \text{ for } T>T_{\rm QCD} 
\end{array}\right.\,.
\end{equation}
 Constant terms in the potential have been left implicit in this expression, as irrelevant for the axion mass and cosmological evolution.
  The coefficient $\alpha$ parametrizes the power-law behaviour at high temperatures. 
Unless otherwise stated, $\alpha = 8$ will be assumed, 
as suggested by the dilute instanton gas approximation (DIGA) 
and in agreement with some lattice computations~\cite{Borsanyi:2016ksw}. 
Note, however, that the  potential in Eq.~(\ref{QCDpotentialT}) differs from the DIGA-motivated cosine potential  commonly assumed for the QCD case:
     our parametrization ensures that $V(\theta_a,T)$ is a continuous function of the temperature not only around the minimum but for any field value $\theta_a$.  
 
\begin{figure}[ht]
\centering
\includegraphics[width=0.65\textwidth]{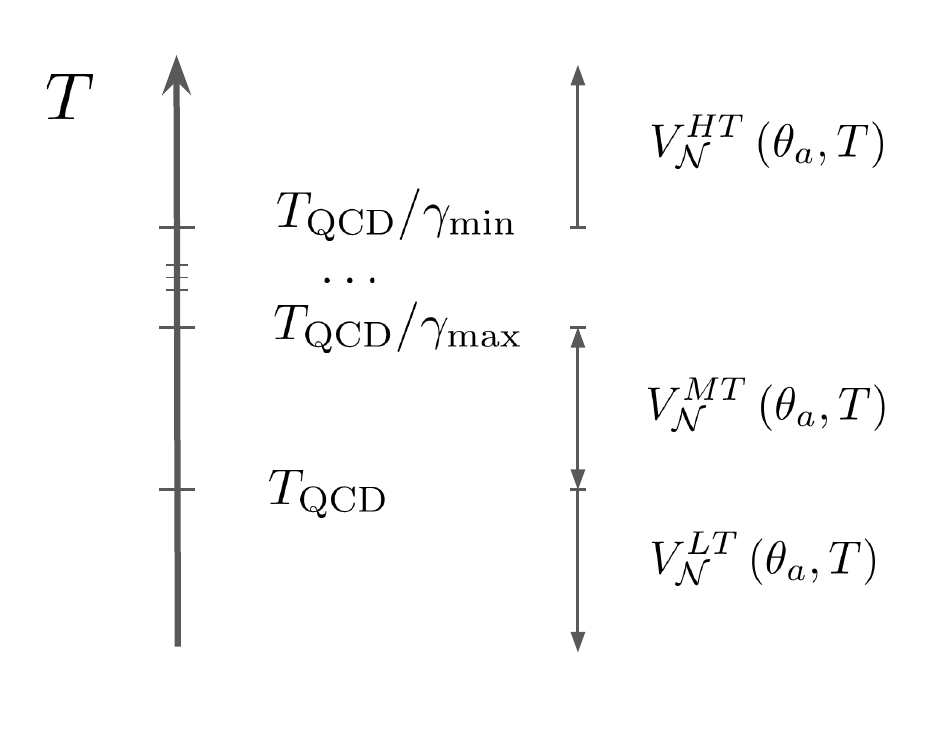} 
\caption{Range of temperatures corresponding to qualitatively different  behaviours of the $Z_\N$ axion potential. The acronyms $HT, MT$ and $LT$ stand for high, medium and low temperatures, respectively.}
\label{fig:Ranges of temperature}       
\end{figure}

The generalization of the ansatz above to the total $Z_\N$ axion potential is simply given by the sum over the $\N$ different potentials,\footnote{This potential depends on all the temperatures of the different mirror worlds, but the simplified notation $V_\N(\theta_a,T, T_1\dots T_{\N-1}) \equiv V_\N(\theta_a,T)$ is used here for simplicity. This notation also reflects the fact that the SM temperature $T$ --being the largest-- is the dominant driver of the Universe evolution. 
}
\begin{equation}
\label{ZNpotentialT}
  V_\N(\theta_a,T)= - \sum_{k=0}^{\N-1}
 m_{\pi}^{2} f_{\pi}^{2}\sqrt{1-\frac{4 m_{u} m_{d}}{\left(m_{u}+m_{d}\right)^{2}} \sin ^{2}\left(\frac{\theta_{a}}{2}+\frac{\pi k}{\mathcal{N}}\right)}\times  h(T_k)\,,
 \end{equation}
  where the power-law temperature suppression is assumed to   be $k$-independent and as given in  Eq.~(\ref{h}), that is,  $\alpha_k =\alpha$ for all $k$. Furthermore, $h(T_k) =  h(b_k\gamma_k T)\simeq  h(\gamma_k T)$, see  \cref{gammalong} and \cref{App: ratios gs grho}.
 Irrespective of the specific $\gamma_{k\ne 0}$ values, three  temperature regimes can be distinguished which are qualitatively different. They 
are determined by the minimum  and maximum value of $\gamma_{k\ne 0}$,
\begin{equation}
\gamma_\text{min}\, \le  \gamma_k\, \le \gamma_\text{max}\,,
\end{equation} 
corresponding respectively to the coolest ($\gamma_\text{min} T$) and the hottest ($\gamma_\text{max} T$) mirror copies of the SM, see Fig~\ref{fig:Ranges of temperature}.

\subsubsection*{Low temperatures $T<T_{\rm QCD}$}
The potential for all $\N$ replicas is well approximated by the --exponentially suppressed-- zero-temperature potential in  Eqs.~(\ref{Eq: fourier potential large N hyper}) and (\ref{maZNLargeN}), 
  \begin{align}
  \label{VlowT}
V^{LT}_{\mathcal{N}}\left(\theta_{a},T\right)\,=\, V_{\mathcal{N}}\left(\theta_{a}\right)\,, 
\end{align}
with minima at $2\pi k/\N$ for $\N$ odd.

\subsubsection*{Medium temperatures $T_{\rm QCD}<T< T_{\rm QCD}/\gamma_{\text{max}}$ } 
In this regime, only the SM contribution to the total potential is temperature suppressed. It is weighed down  by a factor $\left(T_{\rm QCD}/T\right)^\alpha$, while all other contributions are still well approximated by their zero-temperature potential, 
\begin{align}
\label{Eq:Medium Temp potential}
V^{MT}_{\mathcal{N}}\left(\theta_{a},T\right) &\simeq 
\left(\frac{T_{\rm QCD}}{T}\right)^\alpha V(\theta_a)+\sum_{k=1}^{\N-1}V(\theta_a+2\pi k/\N)\nonumber\\
&=  \left[\left(\frac{T_{\rm QCD}}{T}\right)^\alpha-1\right] V(\theta_a) + \sum_{k=0}^{\N-1}V(\theta_a+2\pi k/\N)\nonumber\\
&\simeq  \left[\left(\frac{T_{\rm QCD}}{T}\right)^\alpha-1\right] V(\theta_a)
 -m_{\pi}^{2} f_{\pi}^{2}\frac{1}{\sqrt{\pi}} \sqrt{\frac{1-z}{1+z}} \mathcal{N}^{-1 / 2}(-1)^{\mathcal{N}} z^{\mathcal{N}} \cos \left(\mathcal{N} \theta_{a}\right) \nonumber \\
&
\xrightarrow{\,\,T\gg T_{\rm QCD}\,\,} -V(\theta_a) \, ,
\end{align}
where in the last step the term  $\propto z^\N$ has been neglected compared to the size of 
$V(\theta_a)$. It follows that  the  total axion potential is well approximated  in this regime by {\it minus} the QCD zero-temperature axion potential, that is, the single $k=0$ world contribution with the opposite sign. The potential thus depends only on $f_a$ and  its minimum is located at $\theta_a=\pm\,\pi$. 
  
\subsubsection*{High temperatures $T>T_{\rm QCD}/\gamma_\text{min}$}   
\begin{figure}[ht]
\centering
\includegraphics[width=0.495\textwidth]{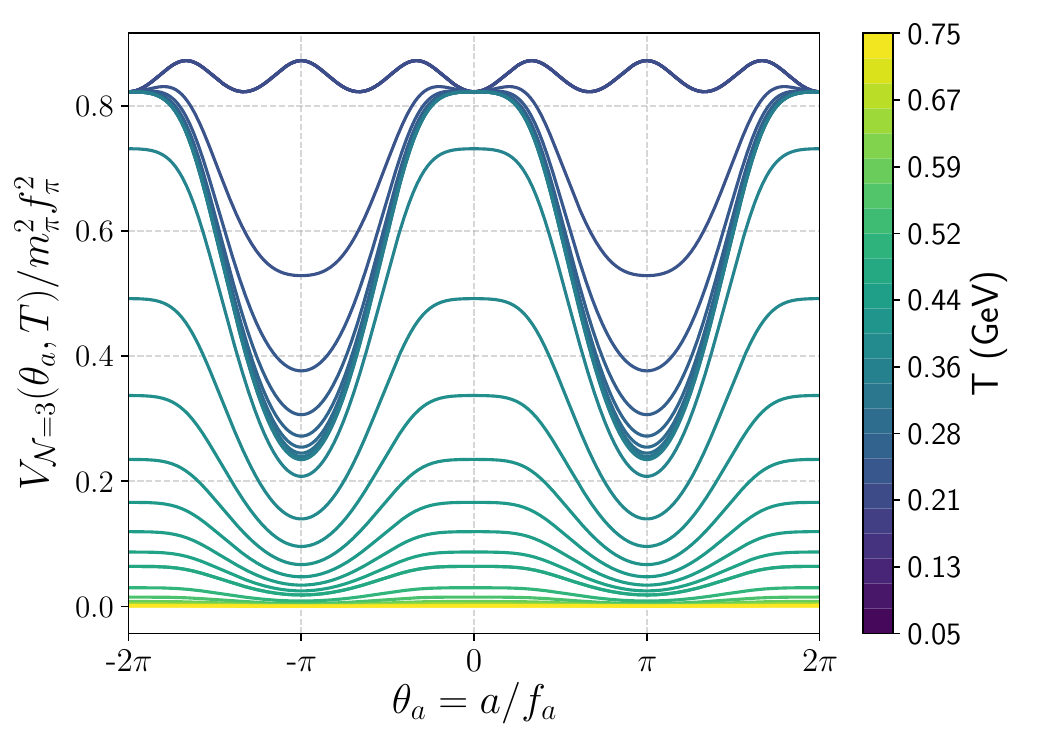} 
 \includegraphics[width=0.495\textwidth]{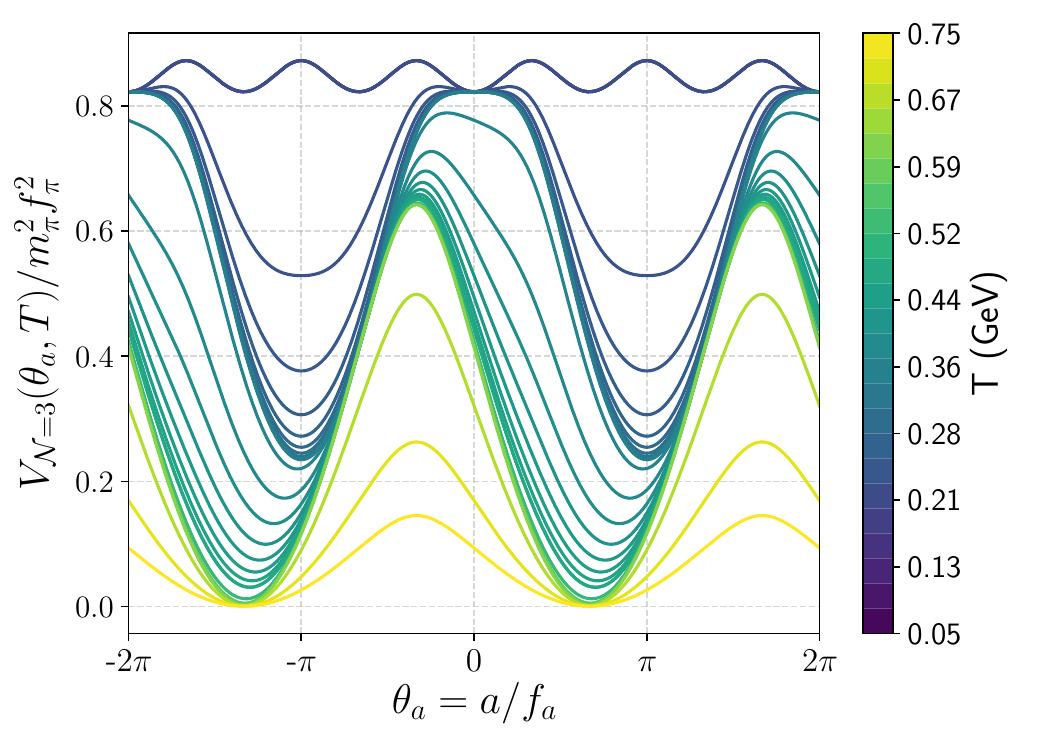} 
\caption{Temperature dependence of the $Z_3$ axion potential. Left panel: all mirror worlds have the same temperature, 
$\gamma_k=\gamma=0.5$. Right panel: mirror worlds have different temperatures, $\gamma_1=0.5,\,\gamma_2=0.25$. 
The low and medium temperature potentials do not depend on $\gamma_{k}$, as it can be appreciated by comparing the two plots.}
\label{fig:Z3temp}       
\end{figure}

The temperature effects are now important for all $\N$ replicas.  In the simplified case with all $\N-1$ mirror worlds having the same temperature 
($T_{k>0}\equiv T'$, $\gamma_{k>0}\equiv\gamma'$, 
$b_{k>0} \equiv b'$) 
\begin{align}
V_{\mathcal{N}}^{H T}\left(\theta_{a}, T\right) &\simeq 
\left(\frac{T_{\rm QCD}}{T}\right)^\alpha \,V(\theta_a)
+\left(\frac{T_{\rm QCD}}{T'}\right)^\alpha\,\,\sum_{k=1}^{\N-1}V(\theta_a+2\pi k/\N)\nonumber \\ 
&\xrightarrow{\,\,\N \gg 1\,\,}-\left(\frac{ T_{\rm QCD}}{T\,\gamma' \,b'}\right)^{\alpha}V\left(\theta_{a}\right)\left[1-(\gamma'\,b')^{\alpha}\right] 
\xrightarrow{\,\,\gamma' \ll 1\,\,} -\left(\frac{ T_{\rm QCD}}{T\,\gamma' \,\,b'}\right)^{\alpha}V\left(\theta_{a}\right) \, ,
\label{Eq: high temp fot equal T}
\end{align}
where in the second step we neglected again the exponentially suppressed $Z_\N$ contribution at $T=0$ 
and  the definition $T' \equiv T \gamma' b'$ was used.  As for the medium temperature regime, the total axion potential 
depends essentially  only on $f_a$ and its minimum is located at $\theta_a=\pm\, \pi$.  In the more general case in which the mirror copies of the SM have different temperatures, only the coldest  world will give a sizeable contribution to the potential, and the minimum will no longer be at 
 $\pm\,\pi$ although typically not too far from it, see Fig.~\ref{fig:Z3temp}.
 
In summary, temperature effects lead to regimes in which the minimum of the total axion potential is displaced to $\theta_a=\pm \, \pi$ or similar large field values. This  high-temperature shift of the minimum parallels that at finite density, previously identified in Refs.~\cite{Hook:2017psm,Huang:2018pbu,ZNCPpaper} and recently employed as well in Ref.~\cite{Brzeminski:2020uhm}. 
This phenomenon is rich in consequences. In particular, it follows that the  axion mass at finite temperature, 
\begin{align}
m_a(T)=\frac{1}{f_a}\sqrt{\frac{d^2V_\N(\theta_a,T)}{d\theta_a^2}\Bigg|_{\rm min}}\,\,,
\end{align}
 is given --in the three temperature regimes discussed above--  by  
 \begin{align}
m_a(T)\simeq \frac{m_{\pi} f_{\pi}}{f_a}  \left\{\begin{array}{lll} \displaystyle
\frac{1}{\sqrt[4]{\pi}} \sqrt[4]{\frac{1-z}{1+z}} \,\,\mathcal{N}^{3 / 4} \, z^{\mathcal{N}/2} 
      &\qquad \text{ for } \ \ \ T<T_{\rm QCD}  \qquad \qquad  \quad \quad \quad \quad \text{LT}\\[12pt]
    \displaystyle 
\frac{\sqrt{z}}{1-z^2} 
    & \qquad \text{ for } \quad T_{\rm QCD}<T< T_{\rm QCD}/\gamma_{\text{max}}\qquad \! \text{MT}\\[12pt]
\displaystyle \frac{\sqrt{z}}{1-z^2}\left(\frac{ T_{\rm QCD}}{\gamma_\text{min}  \,T\, b_{\rm min}}\right)^{\alpha} 
      & \qquad \text{ for } \quad T_{\rm QCD}/\gamma_\text{min}<T\qquad \qquad \quad \ \  \text{HT}\,.
\end{array}\right.
\label{Eq: axion mass temp}
\end{align}
The LT expression for $m_a(T)$ equals that for the zero-temperature mass $m_a$ in Eq.~(\ref{maZNLargeN}), as expected. In contrast, 
 for the intermediate temperature case it corresponds to the tachyonic mass of the QCD axion at the top of its potential,
 \begin{align}
m_{a,\pi}^{\rm QCD}\equiv\frac{1}{f_a}\sqrt{\frac{d^2V(\theta_a)}{d\theta_a^2}}\Bigg|_{\theta_a=\pi}\,= \,\sqrt{\frac{1+z}{1-z}}\,\,m_{a}^{\rm QCD}\,,
\label{mapi}
\end{align} 
which differs from $m_{a}^{\rm QCD}$ by a factor $\sim 1.7$ because the QCD chiral axion potential is not a cosine, and its curvature around maxima and minima are slightly different. 
Thus Eq.~(\ref{Eq: axion mass temp}) can be rewritten as 
 \begin{align}
m_a(T)\simeq \left\{\begin{array}{lll} \displaystyle
m_a &\qquad \text{ for } \ \ \ T<T_{\rm QCD}   \qquad \qquad  \quad \quad \quad \quad  \text{LT}\\[12pt]
\displaystyle m_{a,\pi}^{\rm QCD} 
     & \qquad \text{ for } \quad T_{\rm QCD}<T< T_{\rm QCD}/\gamma_{\text{max}}\qquad \! \text{MT}\\[12pt]
\displaystyle m_{a,\pi}^{\rm QCD}\,\left(\frac{ T_{\rm QCD}}{\gamma_\text{min}  \,T\, b_{\rm min}}\right)^{\alpha} 
    & \qquad \text{ for } \quad T_{\rm QCD}/\gamma_\text{min}<T\qquad \qquad \quad \ \ \text{HT}\,.
\end{array}\right.
\label{Eq: axion mass temp-2}
\end{align}
In summary, at low temperatures the $Z_\N$ axion mass depends both on $f_a$ and $\N$ and 
 is exponentially suppressed by a factor $z^{\N/2}$. In contrast, at 
medium and high temperatures it only depends on $f_a$  in the large $\N$ limit, and  is of the order of the canonical QCD axion  mass  
in vacuum and at high temperature 
respectively.

\subsection {$Z_\N$ axion misalignment}
\label{Sec: ZNaxion misalignemnt}

We analyze here the consequences of the temperature-dependent $Z_\N$ axion potential for  axion DM production. The axion relic density can be estimated from the solution to the equation of motion (EOM) for a classical, non-relativistic and homogeneous scalar field 
$\theta_a$
in an expanding Universe: 
\begin{align}
\ddot{\theta}_a+3 H \dot{\theta}_a+\frac{1}{f_a^2}V'(\theta_a)=0\,,
\end{align}
where $V'(\theta_a)={dV(\theta_a)}/{d\theta_a}$ and the spatial gradients have been neglected. For a field oscillating near its minimum, $V'(\theta_a)\simeq  m_a^2(T)f_a^2\, \theta_a$ is a good approximation and the potential matches that of a damped harmonic oscillator. The behaviour of the solution depends then on the interplay between two variables, the axion mass $m_a(T)$ and the Hubble parameter $H(T)$, which both vary with temperature. 

\begin{figure}[!th]
\centering
\begin{subfigure}[h]{0.495\textwidth}
 \includegraphics[width=\textwidth]{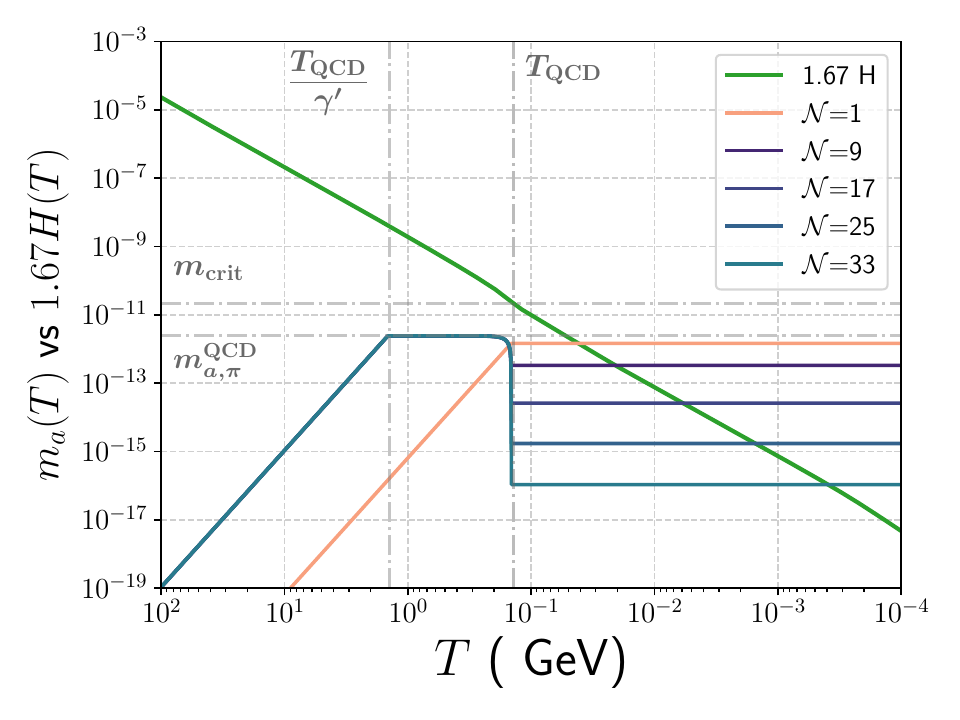} 
 \caption{$f_a=4\times 10^{18}\,\text{GeV} >f_a^{\rm crit}$: Simple ALP}
 \end{subfigure} 
 \begin{subfigure}[h]{0.495\textwidth}
 \includegraphics[width=\textwidth]{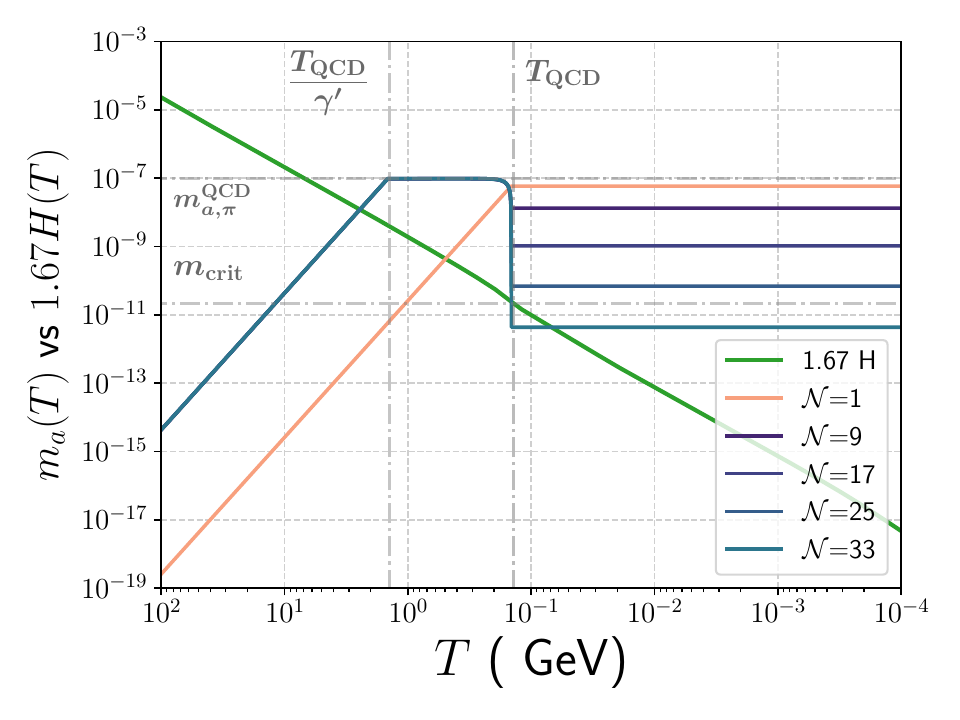} 
 \caption{$f_a= 1\times 10^{14}\,\text{GeV}<f_a^{\rm crit}$: Trapped mis.}
 \end{subfigure}
  \caption{Axion mass vs.~Hubble temperature dependence. 
Left panel: ``simple ALP'' misalignment regime
(i.e.~scalar with constant mass) corresponding to one stage of oscillations. Right panel: trapped misalignment regime with two stages of oscillations. 
Note the degeneracy in $\N$ of the trajectories in the medium and high-temperature regimes. The value $\gamma_{k>0} \equiv \gamma' = 1/10$ has been used. }
\label{fig:tempaxionmass}       
\end{figure}

We discuss next the evolution of the axion mass as a function of $T$, $f_a$ and $\N$. 
 As the Universe cools down, the axion mass  progressively grows from its initial vanishingly small value at high temperatures, until it suddenly drops down around  $T_{\rm QCD}$ (feeling the $z^\N$ exponential suppression of the potential) and reaches then a constant value. This is shown in  \cref{fig:tempaxionmass}, where the $Z_\N$ trajectories are depicted in blue for different values 
 of $\N$. The figure illustrates the $\N$-independence of $m_a(T)$ at high and medium temperatures 
 (in the large $\N$ limit), while it becomes $\N$-dependent at low-temperature, see also Eq.~(\ref{Eq: axion mass temp}) and \cref{fig:Z3temp}. 
  At some point in that trajectory, the mass overcomes the Hubble friction --depicted in {\color{HubbleGreen} \bf green}--  and the axion starts to oscillate. 

  There are two qualitatively distinct regimes, depending on whether this crossing happens after or before the axion mass abruptly drops at $T_{\rm QCD}$ (as illustrated 
  in \cref{fig:tempaxionmass}).  Let us denote by $f_a^{\rm crit}$ the value of $f_a$  
  for which the canonical QCD axion would start to oscillate precisely at $T_{\rm QCD}$, 
\begin{equation}
f_a^{\rm crit}\equiv \frac{m_\pi f_\pi}{m_{\rm crit}}\frac{\sqrt{z}}{1+z}\simeq2\times 10^{17}\, \text{GeV}, \quad \text{where} \quad m_{\rm crit}=1.67\,H(T_{\rm QCD})\simeq 2.9\times 10^{-11}\, \text{eV}\,.
\label{Eq:mcrit and fcrit}
\end{equation} 
Whenever $m_a> m_{\rm crit}$ ($m_a< m_{\rm crit}$) Hubble crossing takes place just (more than) once in the cosmic trajectory. The factor $1.67$ for 
 $m_{\rm crit}$  stems from our criteria for the onset oscillations that better fits the numerical solution to the EOM, as argued further below.
Strictly speaking, the value of $f_a^{\rm crit}$  for the $Z_\N$ axion is slightly larger than that in Eq.~(\ref{Eq:mcrit and fcrit}), 
see Eq.~(\ref{mapi}); this small difference will be taken into account in the numerical results and figures, but disregarded in the discussion below for the different regimes of oscillation. In \cref{fig:tempaxionmass}, $f_a=f_a^{\rm crit}$ corresponds to the case in which the ({\color{HubbleGreen} \bf green}) Hubble line would cross  the   canonical QCD axion trajectory (i.e.~$\N=1$, {\color{orangeQCD} \bf orange}) exactly at the flattening point of the latter, 
 i.e.~$\sim T_{\rm QCD}$.

\subsubsection{One stage of oscillations: simple ALP regime} 
For {\boldmath $f_a>f_a^{\rm \bf{crit}}$} the $T=0$ potential is already developed when the axion starts to oscillate, 
 see left panel of \cref{fig:tempaxionmass}.  
 Oscillations only take place  around the true minimum $\theta_a=0$, and the  relic density corresponds to that of a ``simple ALP'' with constant mass. 
 This single regime of oscillations with constant mass  applies as well to the 
standard QCD axion scenario (i.e.~$\N=1$ in \cref{fig:tempaxionmass}) for large enough $f_a$.
The evolution of the axion field is straightforward and goes as follows:
\paragraph*{$T>T_{1}$.}  At high temperatures the axion field is frozen at a given and arbitrary initial misalignment angle $\theta_1$ due to Hubble friction, $H\gg m_a(T)$.  As the Universe cools down the Hubble parameter diminishes, until the friction term intercepts the $m_a(T)$ trajectory  at a temperature $T_1$ such that  
\begin{equation}
1.67H(T_1)=m_a(T_1)\equiv m_1\, .
\label{T1}
\end{equation}
Since $T_1<T_{\rm QCD}$, the true vacuum potential has already developed  when the axion field starts to oscillate. 
 In consequence,  its mass at $T_1$ is already  the zero-temperature mass:
\begin{equation}
m_1=m_a\,, \qquad \text{with} \qquad \theta_a(T_1)=\theta_1\,,\quad \dot{\theta}_1=0\,.
\label{param-onestage}
\end{equation}
\paragraph*{$T_{1}>T$.} 

The axion fields keeps oscillating around $\theta_a=0$ until nowadays and its mass is almost constant, $\dot{m_a} \simeq 0$. It is  adequate to apply the WKB approximation provided  $m_a\gg\big\{\dot {m}_{a}/{m}_{a},\dot H/H\big\}$, i.e.~provided  the oscillations are much faster than both the expansion rate of the Universe and the rate at which  the mass changes, which is the case here. The adiabatic approximation predicts the existence of a conserved quantity~\cite{Preskill:1982cy,Abbott:1982af,Dine:1982ah}, an adiabatic invariant $N_a$ that can be interpreted as the comoving number of non-relativistic axion quanta, defined for a generic axion mass $m_a$ as 
\begin{equation}
N_a\equiv\frac{\rho_a a^3}{m_a}= {\rm const.}\, , \qquad 
\text{with}\qquad {\rho_{a}} \equiv\frac{1}{2} {\dot \theta}_a^2 {f_a^2}+\frac{1}{2} m_{a}^2{f_a^2} \theta_a^2\,,
\label{Na}
\end{equation}
where $a$ denotes the scale factor and $\rho_a$ the axion energy density.   For the regime under discussion, 
 it follows from $N_a$  conservation between $T_1$ and today that the current energy density $\rho_{a,\,0}$ can be expressed as
\begin{align}
\rho_{a,\,0}=m_{a} \frac{N_a}{a_{0}^{3}} \simeq \frac{1}{2} \,m_{a}^2 \, (\theta_{1}\,f_a)^2\,\left(\frac{a_{1}}{a_{0}}\right)^{3}\,,
\label{Eq: relic density simple ALP a}
\end{align}
where $a_0$ and $a_1$ denote respectively the value of the scale factor today and at the onset of oscillations, $a_1\equiv a(T_1)$.\footnote{Whenever the initial misalingment angle is large, the relic density in \cref{Eq: relic density simple ALP a} needs to be corrected by the so-called anharmonicity factor, see \cref{sec:anharmonicity_function}.} 
 The latter can be expressed in turn in terms of the axion mass using Friedmann equations and the conservation of entropy, $a_1\propto 1/\sqrt{m_a M_{\rm Pl}}$,  allowing us to obtain the ratio of the axionic relic density to the current  DM abundance $\rho_{\rm DM}$ as a function of $m_a$, $\theta_1$ and $f_a$,\footnote{The slight discrepancy with respect to the result in Ref.~\cite{Arias:2012az} is due to the use of the updated value of $\rho_{\rm DM}\simeq 1.26\, {\mathrm{keV}}/{\mathrm{cm}^{3}}$ from Planck 2018 data~\cite{Aghanim:2018eyx} and a different choice for the onset of the oscillations that better fits the numerical result 
($1.67 \, H=m_a$ instead of $3H=m_a$, see e.g.~Refs.~\cite{AlonsoAlvarez:2019cgw,Marsh:2015xka}).}

 \begin{align}
  \frac{\rho_{a,0}}{\rho_{\rm DM}} \simeq 5.7  \sqrt{\frac{m_{a}}{\mathrm{eV}}} \left(\frac{\theta_{1}\,f_a}{ 10^{12}\, \mathrm{GeV}}\right)^{2} \mathcal{F}(T_{1})\,,
  \label{Eq:simple ALP relic dansity ratio CMB}
  \end{align}
 where $\mathcal{F}(T_{1}) \equiv\left(g_{*}(T_{1}) / 3.38\right)^{\frac{3}{4}}\left(g_{s}(T_{1}) / 3.93\right)^{-1}$ is an $\mathcal{O}(1)$ factor and $\rho_{\rm DM}\simeq 1.26\, {\mathrm{keV}}/{\mathrm{cm}^{3}}$ from Planck 2018 data~\cite{Aghanim:2018eyx}.  
   For the $Z_\N$ scenario under discussion, it follows from \cref{maZNLargeN} that this ratio can be rewritten  in the large $\N$ limit  as 
  \begin{align}
  \frac{\rho_{a,0}}{\rho_{\rm DM}}  \simeq 0.48 \, \N^{3/8}{z^{\N/4}}\,  \theta_{1}^2\, \left(\frac{f_a}{ 10^{13} \mathrm{GeV}}\right)^{3/2} \mathcal{F}(T_{1})\,.
  \label{eq:rhoa0simpleALP}
  \end{align}
The initial axion misalignment $\theta_1$ can {\it a priori} take any value in the $[-\pi,\pi)$ interval and the axion field will roll down to the closest minimum\footnote{Note that the relevant axion field displacement that should enter in \eq{eq:rhoa0simpleALP} corresponds to the field distance to the closest minimum and it is only when the closest minimum is $\theta_a=0$ that the field distance coincides with $\theta_1$.} among the $\N$
 {possibilities}. Thus it is with probability  $1/\N$ that the final $\theta$-parameter will correspond to the CP-conserving point $\theta=0$.
 In other words, in order for the simple ALP regime to solve the strong CP problem and also account for DM  it is necessary and sufficient that the initial misalignment angle lies in the interval $\theta_1\in[-\pi/\N,\pi/\N)$ within the simple ALP regime, 
for any $f_a>2\times 10^{17}\, \text{GeV}$.

 Eq.~(\ref{Eq: relic density simple ALP a})	also indicates that  the simple ALP solutions which account for the relic DM obey (see \cref{fig:Sketch}) 
 \begin{align}
  m_af_a^4 \propto  {\rm const.}\,,
  \label{mafaALP}
  \end{align}
to be compared with the $Z_\N$ axion mass relation, that for a given $\N$ predicts $ m_af_a \propto  {\rm const.}$ This behaviour is  depicted in the $\{m_a,1/f_a\}$ plane in Fig.~\ref{fig:axionDM} by continuous superimposed  lines  (for different values of $\theta_1$), together with some  experimentally excluded regions as a reference (while the complete set of present and projected sensitivity regions is depicted in \cref{fig:axionEDM}).  It follows from Fig.~\ref{fig:axionDM}  that the  simple ALP $Z_\N$ solutions  can solve both the strong CP problem and account for DM mass down to the region of fuzzy DM ($m_a \sim 10^{-22}$ eV) 
and for any value of $\N\ge3$.

\subsubsection{Two stages of oscillation: trapped misalignment}
\begin{figure}[!ht]
           \centering
           \includegraphics[width=0.99\textwidth]{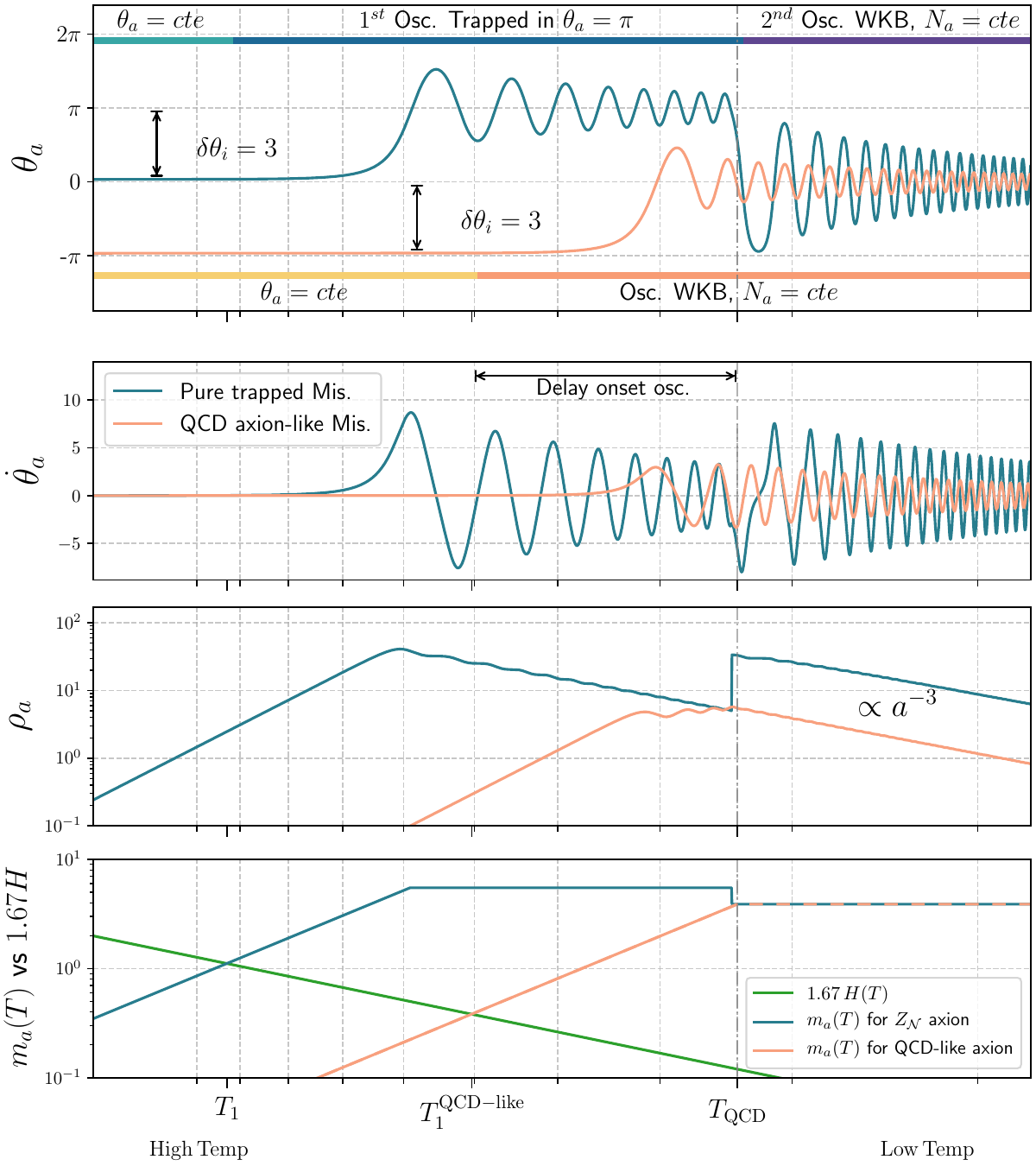} 
           \caption{Axion evolution for the {\bf Pure Trapped misalingment} vs.~that for a canonical QCD axion, illustrating the relic density enhancement for the same final zero-temperature mass $m_a$.}
           \label{fig:Trapped VS wispy Evolution}       
           \end{figure}

\begin{figure}[!h]
          \centering
          \includegraphics[width=0.99\textwidth]{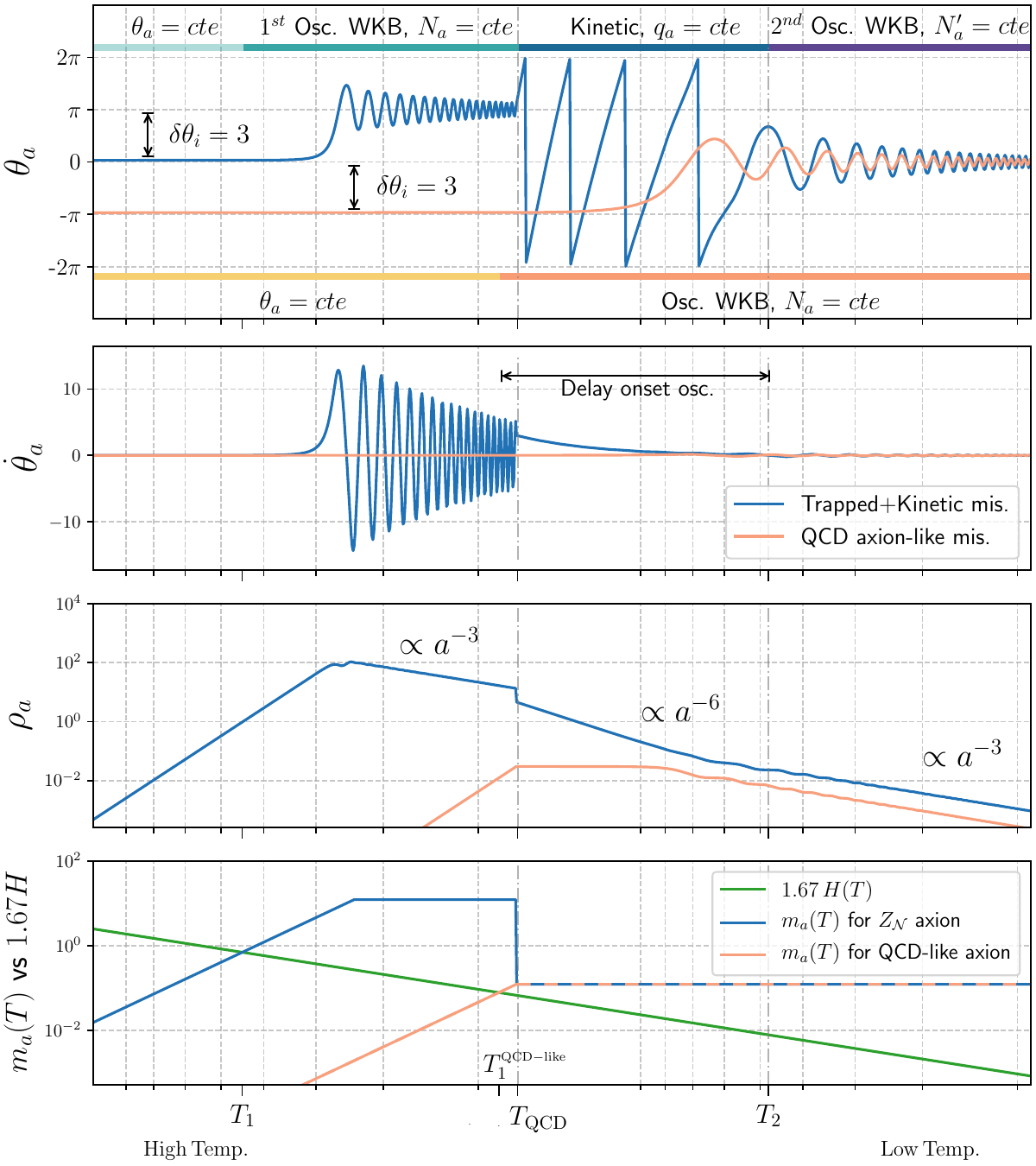} 
          \caption{Axion evolution for the {\bf Trapped+Kinetic misalignment} vs.~that for a canonical QCD axion, illustrating the strong enhancement of the relic density enhancement for the same final zero-temperature mass $m_a$.  }
          \label{fig:Trapped+Kin Evolution}       
          \end{figure}

 For  {\boldmath $f_a<f_a^{\rm crit}$},  two different potential minima develop during the Universe evolution from high to low temperatures, see \cref{fig:tempaxionmass} (right panel). In a nutshell:
\begin{itemize}
\item A first phase of oscillations takes place around $\theta_a=\pi$ at temperatures above $T_{\rm QCD}$,    
because temperature effects are important then and the axion mass is unsuppressed.
\item At $T_{\rm QCD}$ the true minimum $\theta_a=0$ develops (while $\pi$ becomes a maximum). The axion mass suddenly becomes exponentially suppressed due to the $Z_\N$ symmetry. Oscillations around $\theta_a=0$ will start whenever the kinetic energy becomes smaller than the height of the potential barrier.
\end{itemize} 
These two stages  are thus separated by a drastic (non-adiabatic) modification of the potential. 
This novel axion DM production mechanism is named
 \emph{trapped misalignment}. Its cosmic evolution is illustrated in Figs.~\ref{fig:Trapped VS wispy Evolution} and \ref{fig:Trapped+Kin Evolution} for  toy-model trapped trajectories ({\color{BlueKin} \bf blue}), versus that of a QCD-like axion ({\color{orangeQCD} \bf orange}) with the same zero temperature mass. 
Let us expatiate next on the evolution details.

\paragraph*{$T>T_{1}$.} In this period, the behaviour is alike to that for a simple ALP, see Eq.~(\ref{T1}).  Before Hubble crossing at $T_{1}$, 
 the axion field is \emph{frozen}  and remains constant at an arbitrary initial misalignment angle $\theta_1$. 
\paragraph*{$T_{1}> T>T_{\rm QCD}$.}  The finite temperature effective potential is relevant in this range,  see Eqs.~(\ref{Eq:Medium Temp potential}) and (\ref{Eq: high temp fot equal T}), and the initial axion velocity is $\dot\theta_1=0$. 
  Thus, the axion  ``is trapped'' in this \emph{first stage of oscillations} around  $\theta_{a}\sim\pi$  from $T_1$ until $T_{\rm QCD}$.   Its mass is unsuppressed due to the explicit $Z_\N$ breaking by the thermal background:  $m_a(T)$  raises  progressively until it stabilizes at $\sim m_a^{\rm{QCD}}$.  

Trapping   causes a {\it delay of the start of oscillations around the true minimum $\theta_a=0$,} as compared to $T_1$ for the canonical QCD axion and the simple ALP scenarios. The consequence is an {\it enhancement of the DM density}, as the dilution time is thus shortened.  

In \cref{fig:Trapped VS wispy Evolution,fig:Trapped+Kin Evolution} the evolution for a QCD-like axion (i.e.~$\N=1$, {\color{orangeQCD} \bf orange}) is shown to  oscillate around   $\theta_a=0$ since  well before $T_{\rm QCD}$.  In contrast, for $\N\ge3$ ({\color{BlueTrapped} \bf blue}) the oscillations around $\theta=\pi$ take place from $T_1$ down to $T_{\rm QCD}$. The crucial impact of the trapped stage is thus to delay the onset of  oscillations around the true minimum, and to set the initial value of $\theta_a$ and $\dot{\theta_a}$ at $T_{\rm QCD}$, 
  \begin{equation}
 \theta_{\rm tr}\equiv\theta_a(T_{\rm QCD})\sim\pi,\,\quad \dot\theta_{\rm tr}\equiv\dot\theta_a(T_{\rm QCD})\,.
 \label{trapped-param}
 \end{equation}
 It is not possible to predict the exact value of $\dot\theta_{\rm tr}$.  However, an order of magnitude estimate  stems  from   the energy density when the temperature approaches $T_{\rm QCD}$,   applying  $N_a$ conservation to the adiabatic interval  $[T_1,T_{\rm QCD})$,
   \begin{align}
       \rho_{a,\,{\rm QCD}}\equiv \rho_{a}(T_{\rm QCD})= m_{a,\pi}^{\rm QCD} \frac{N_a}{a_{\rm QCD}^3}= \frac{1}{2}m_1 m_{a,\pi}^{\rm QCD} \left(\frac{a_1}{a_{\rm QCD}}\right)^3 (\theta_{1}-\pi)^ 2f_a^2 \, , 
      \end{align}
      which shows that the precise value of $\rho_{a,\,{\rm QCD}}$ depends on the arbitrary initial misalignment angle with respect to the high-temperature minimum in $\pi$: $(\theta_1-\pi)$.  This dependence is inherited by the mean velocity at the end of this period, which  
       follows  from the equality 
       of the mean kinetic and potential energies for a harmonic oscillator,
      \begin{align}
      \sqrt{\langle \dot \theta^2_{\rm tr}\rangle}= \frac{1}{\sqrt{2}}\sqrt{m_1 m_{a,\pi}^{\rm QCD}} \left(\frac{a_1}{a_{\rm QCD}}\right)^{3/2} |\theta_{1}-\pi| \,.
      \label{Eq Mean theta dot}
      \end{align}

\paragraph*{$T=T_{\rm QCD}$.}  At this point,  the low-temperature  potential for the  $Z_\N$  axion scenario develops (see Eqs.~(\ref{Eq: fourier potential large N hyper}),  (\ref{maZNLargeN}) and (\ref{VlowT})), and thus $\theta_a=\pi$ becomes a maximum. The abrupt exponential suppression of the axion mass
  was shown in \cref{fig:tempaxionmass}  for different values of $\N$. It illustrates that the two stages of oscillation  are  separated by a sudden, inherently non-adiabatic, modification of the potential as $\dot m_a/m_a\gg m_a$. 
 The energy density of the axion just after the transition at $T_{\rm QCD}$ reads
  \begin{align}
   {\rho_{a,{\rm tr}}} =\frac{1}{2} {\dot \theta}_{\rm tr}^2 {f_a^2}+ \frac{m_{a}^2{f_a^2}}{\N^2}\big[1- \cos(\N\theta_{tr})\big]\,\simeq \frac{1}{2} {\dot \theta}_{\rm tr}^2 {f_a^2}+ 2 \frac{m_{a}^2{f_a^2}}{\N^2}\,,
   \label{rho-tr}
   \end{align}
   that is, the total energy density at this point is larger or equal than the maximum potential energy (the height of the barrier $ 2 {m_{a}^2{f_a^2}}/{\N^2}$).
   What happens next depends crucially on the  value of $\dot\theta_{\rm tr}$, which acts as initial condition for the subsequent period. 
   It determines at which temperature the \emph{second stage of oscillations} begins:
   \begin{itemize}
   \item For very small axion velocity,  
    $\dot \theta_{\rm tr}\ll 2 m_a/\N$, oscillations around one minimum may start closely after $T_{\rm QCD}$. This case will be denoted below  ``pure trapped misalignment''. 
   \item For   large enough $\dot\theta_{\rm tr}$ so that the kinetic energy  dominates ${\rho_{a,{\rm tr}}}$,  $\dot \theta_{\rm tr}\gg 2m_a/\N$, the axion field will roll  over the top of the  barriers for a long time before its oscillations start at a temperature sensibly lower than $T_{\rm QCD}$. This case will be denoted as ``trapped+kinetic misalignment''. 
     \end{itemize}

\paragraph{Pure trapped misalignment: $\dot\theta_{\rm tr}\sim 0$} 
\paragraph*{$T_{\rm QCD}> T$.}   Were the axion velocity  exactly zero at $T_{\rm QCD}$, $\dot\theta_{\rm tr}= 0$, the setup would not be viable: 
 the axion would simply roll to the closest minimum $\theta_a=\frac{\pi}{\N}(\N-1)$. This  is precluded by the experimental limit on the nEDM that implies $\theta \lesssim 10^{-10}$. 
 
 Nevertheless, $\dot\theta_{\rm tr}$ may be very small but not zero. Even a tiny kinetic energy may suffice to allow the axion to go over some potential barriers, because the misalignment set by the trapping corresponds to a  maximum of the zero temperature potential. In Sect.~\ref{sec:CPconsminimum} the minimum kinetic energy necessary to roll over $\mathcal{O}(\N)$ potential barriers\footnote{Strictly speaking, about $ \N/2$ barriers must be overflown to approach the $\theta_a=0$ minimum, but we will obviate this finesse as it  does not change the order of magnitude estimates intended here.} will be estimated and shown to be attained in most of the parameter space.  The axion field will  then end up in the CP-conserving minimum $\theta_a=0$ with $1/\N$ probability. 
 
An estimation of the time required for the axion to go from one maximum to the next is  $\delta t\sim 2\pi/m_a$. To overfly only  $\sim \N$ barriers a very short time is required  because in this regime $m_a \ll H$, i.e.~the difference between $T_{\rm QCD}$ and the temperature at the onset of oscillations around the true minimum   is negligible. In conclusion,  for $\dot \theta_{\rm tr}\ll 2 m_a/\N$  the kinetic energy is not large enough to extend  beyond $\sim T_{\rm QCD}$ the delay time accumulated during the trapped period (see \cref{rho-tr}).  The value of $\theta_a$ when it starts to oscillate from the top of the barrier closest to $\theta_a=0$ is 
 \begin{equation}
 \theta_2\sim \frac{\pi}{\N}\,.
 \label{closest}
 \end{equation}
The evolution is illustrated in \cref{fig:Trapped VS wispy Evolution}, which depicts the numerical solution for a toy example  exhibiting pure trapped misalignment ({\color{BlueTrapped} \bf blue}) and  compares it with that for  a QCD-like axion ({\color{orangeQCD} \bf orange}) with the same zero temperature mass. The figure reflects the enhancement of the relic DM density in the trapped case. The latter can be easily computed analogously to that for the usual misalignment mechanism.   After the transition at $T_{\rm QCD}$, the adiabatic approximation is valid again since  
$\dot m_a\simeq 0$, $m_a\gg H$. From the $N_a$ conservation law --Eq.(\ref{Na}), and the initial conditions at the onset of the second stage of oscillations 
 {$\big\{\theta_{\rm 2}\simeq\pi/\N,\,\dot\theta_{\rm 2}\simeq 0\big\}$},  it follows that  the current axionic relic density $\rho_{a,0}$  is given in the pure trapped case by
         \begin{align}
        \Aboxed{\rho_{a,0}\big|_{\rm tr}= m_a \frac{N_a}{a_0^3}=\frac{1}{2}\,m_a^2\, \left(\frac{\pi }{\N}f_a\right)^2 \,\left(\frac{a_{\rm QCD}}{a_0}\right)^3 \,, }
        \label{Eq: pure trapped relic density}
        \end{align}
        where $a_{\rm QCD}\equiv a(T_{\rm QCD})$.  For a given final vacuum mass $m_a$, it  demonstrates an enhancement of the relic density in the scenario with a  $Z_\N$ trapped axion vs.~that with a simple ALP, for two main reasons: i)  the relative scale dependence is larger by a factor 
$({a_{\rm QCD}}/{a_1})^3$ and does not depend on the axion mass;  ii)  the initial misalignment angle is fixed to $\theta_{\rm tr}\sim \pi/\N$, while for the simple ALP it is arbitrary within the interval $ [-\pi/\N,\pi/\N)$. Note that the current relic density in the pure trapped scenario becomes independent of the initial misalignment $\theta_1$, unlike the case of the simple ALP  and QCD-like pre-inflationary scenarios.
          
For the $Z_\N$ reduced-mass axion under discussion, the product $m_a f_a$ --and thus $\rho_{a,0}$ in Eq.~(\ref{Eq: pure trapped relic density})-- is a function of just one unknown: $\N$, see $m_a$ in Eqs.~(\ref{maZNLargeN}) and (\ref{Eq: axion mass temp}).
Substituting this dependence in the ratio 
  \begin{align}
  \frac{\rho_{a,0}}{\rho_{\rm DM}}\bigg|_{\rm tr} \simeq 5.6  \sqrt{\frac{m_{a}}{\mathrm{eV}}} \left(\frac{m_a}{m_{\rm crit}}\right)^{3/2} \left(\frac{\pi}{\N}\,\frac{f_a}{ 10^{12}\, \mathrm{GeV}}\right)^{2} \mathcal{F}(T_{\rm QCD})\,,
  \label{Eq:simple ALP relic dansity ratio CMB}
  \end{align}
  it follows that  
\begin{align}
\frac{\rho_{a,0}}{\rho_{\rm DM}}\bigg|_{\rm tr}\sim 1.5\times 10^{7}\,\frac{z^{\N}}{\sqrt{\N}}\,.
\end{align}
Thus in the pure trapped regime the value of $\N$ compatible  with simultaneously solving the strong CP problem and accounting for the ensemble of the DM density  is
       \begin{equation}
     \N\sim 21\,.
     \label{N21}
          \end{equation}
   This line of constant relic density  is highlighted in Fig.~\ref{fig:axionDM} ({\color{PurpleTrapped} \bf purple})  for relatively small $f_a$ values. Indeed, Eq.~(\ref{Eq: pure trapped relic density}) predicts that the relic density solution in the $\{m_a,f_a\}$ plane has to be a line parallel to those defining the canonical QCD and $Z_\N$ axion bands, because in all these cases the parametric dependence  is $m_af_a= {\rm const.}$ The estimate in Eq.~(\ref{N21}) disregards anharmonicity corrections 
and in order to qualitatively account for the latter ones 
the axion DM region will be represented as a band in the figures of Sect.~\ref{sec:pheno}.\footnote{The size of this correction  will be different in general for the pure trapped case and the simple ALP regime since in the former $H\ll m_a$, see \cref{sec:anharmonicity_function}.}

\paragraph{Trapped+kinetic misalignment: sizeable $\dot\theta_{\rm tr}$}

 \paragraph*{$T_{\rm QCD}> T>T_{2}$.} 

If the axion velocity at $ T_{\rm QCD}$ is large enough,  $\dot \theta_a\gg 2m_a/\N$,\footnote{Equivalently,  if the axion kinetic energy at the end of the trapped period  $\dot\theta_{\rm tr}^2 f_a^2/2$  is much larger than the maximum potential energy $2m_a^2 f_a^2/\N^2$.} the axion field has enough kinetic energy to roll  many times over the barriers before it starts to oscillate around some minimum, triggering the kinetic misalignment mechanism. The onset of oscillations  is thus  delayed until much later than $ T_{\rm QCD}$.

In this regime, the WKB (adiabatic) approximation is not valid anymore and the axion energy density is diluted as $\propto a^{-6}$ due to the expansion of the Universe (as illustrated in Fig.~\ref{fig:Trapped+Kin Evolution}). This can be shown by noting that $\dot \theta_a f_a^2$ is the Noether charge density associated to the PQ symmetry \cite{Co:2019jts} and therefore it decays as $\dot \theta_a \propto a^{-3}$ as long as the  axion mass can be 
neglected.\footnote{Alternatively, the dilution of the axion velocity can be obtained 
from the axion EOM in the massless limit: 
$\ddot{\theta}_a+3 H \dot{\theta}_a=0\,\implies \dot \theta_a a^3={\rm const}$.} In consequence, a comoving PQ charge $q_{\rm kin}$ can be defined which is a conserved quantity,   
\begin{align}
      q_{\rm kin}=\dot \theta_a a^3={\rm const} \, .
      \label{Eq: q_kin}
      \end{align}
       This regime of rapidly decreasing energy density ends up at a temperature $T_2$ at which the kinetic energy becomes of the order of the height of the barrier, and thus the axion can no longer overcome it, 
  \begin{align}
       \frac{1}{2}\dot \theta^2_{a}(T_2)f_a^2=2\,\frac{m_a^2 f_a^2}{\N^2} \quad \implies \quad  \dot \theta_{a,\,2} \equiv \dot \theta_{a}(T_2)=2\, \frac{m_a}{\N}\,.
       \label{Eq: condition stop kin}
       \end{align}
At this point  the axion will  start a second stage of oscillations, see Fig.~\ref{fig:Trapped+Kin Evolution}.  Enforcing the conservation of $q_{\rm kin}$ from $T_{\rm QCD}$ until $T_2$, and using Eq.~(\ref{Eq: condition stop kin}), it follows that the scale factor at   the onset of the second stage of oscillations 
 $ a_2\equiv a(T_2)$ can be expressed as 
       \begin{align}
       q_{\rm kin}=
       \dot \theta_{\rm tr} \,a_{{\rm QCD}}^3
       =2\, \frac{m_a}{\N} a_{2}^3 \quad \implies \quad 
       a_2= \left(\frac{ \N\,\dot \theta_{\rm tr}}{2\,m_a}\right)^{1/3} a_{{\rm QCD}} \,.
      \label{Eq: a2 onset true osc}
       \end{align}

 \paragraph*{$T_{2}> T$.}
The axion finally starts the second stage of oscillations around one of the $\N$ zero-temperature minima at random. 
 Thus, the $1/\N$ probability of the $Z_\N$ axion scenario~\cite{Hook:2018jle,ZNCPpaper} to solve the strong CP problem (i.e.~the final $\theta$-parameter to corresponding the CP-conserving point $\theta=0$)  will be maintained when accounting for DM.

 In order to determine the axion relic density today, the adiabatic approximation is again appropriate after $T_2$.  The comoving number of axions $N_a'$ is constant during this final period (and different from that during the first oscillation period $N_a$). Using $q_{\rm kin}$    in Eq.~(\ref{Eq: a2 onset true osc}) and  the mean axion velocity at 
       $T_{\rm QCD}$ in \cref{Eq Mean theta dot},  $N_a'$ is well approximated by \cite{Co:2019jts}
        \begin{align}
         N'_a= \frac{\rho_{a,2} a_2^3}{m_a}=2\,\frac{m_a f_a^2 a_2^3}{\N^2}
         = \frac{\dot \theta_{\rm tr}}{\N}\,f_a^2\,a_{\rm QCD}^3 \,.
         \end{align} 
Finally the relic density today, 
\begin{align}
       \rho_{a,0}\big|_{\rm tr+kin}\simeq\,\frac{m_a}{a_0^3}N_a'\,, 
       \label{Eq:relic density a trapped+kinetic-aprox}
       \end{align}
can be written as 
       \begin{align}
       \Aboxed{\rho_{a,0}\big|_{\rm tr+kin}= 
       C\,\frac{m_a\sqrt{m_1 m_{a,\pi}^{\rm QCD}}}{\sqrt{2}\,\N} \left(\frac{\sqrt{a_1 a_{\rm QCD}}}{a_{0}}\right)^{3} |\theta_{1}-\pi|\, f_a^2 \,, } 
       \label{Eq:relic density a trapped+kinetic}
       \end{align}
       where $C$ denotes a deviation from the analytical --exactly adiabatic and harmonic-- estimations above, to be computed numerically. For our setup, we find numerically $C\simeq 2$, which confirms the result first obtained in Ref.~\cite{Co:2019jts}; this correction will be included in all figures below for the kinetic+trapped mechanism.

     The comparison in  \cref{fig:Trapped+Kin Evolution} between the trapped+kinetic evolution ({\color{BlueKin} \bf blue}) with that for  a QCD-like axion ({\color{orangeQCD} \bf orange}) shows how the delay produced by the kinetic-dominated period results in an enhancement of the current axionic relic density, which is even stronger  than in the pure trapped scenario. The period of kinetic misalignment has been shortened in the figure for illustration purposes.
  
         The comparison of the resulting axionic relic density  with that for the pure trapped scenario in \cref{Eq: pure trapped relic density} shows two competing effects. In fact, for the 
         kinetic+trapped scenario:
 \begin{itemize}
 \item the dilution factor is smaller, $(\sqrt{a_1 a_{\rm QCD}}/{a_{0}})^{3}$ vs.~$(a_{\rm QCD}/{a_{0}})^{3}$;
 \item  the relevant mass scale is larger, $\sqrt{m_1 m_{a,\pi}^{\rm QCD}}$ vs.~$m_a$. 
 \end{itemize}
Hence, it will depend on the specific point of the parameter space whether the kinetic+trapped or the pure trapped mechanism gives the dominant contribution (see \cref{subsec:parametrics_of_the_differnt_production_mechanisms}). 

The ratio of the predicted axion relic density to the observed DM density can be obtained following 
same the procedure applied for the simple ALP regime, yielding
          \begin{align}
       \frac{\rho_{a,0}}{\rho_{\rm DM}}\bigg|_{\rm tr+kin} \simeq 7.9  
       \sqrt{\frac{m_{a}}{\mathrm{eV}}} \,
       \frac{\sqrt{m_{a}\,m_{a,\pi}^{\rm QCD}}}{\big(m_{\rm crit}^3 m_1\big)^{1/4}} 
       \left[\frac{\,f_a}{ 10^{12}\, \mathrm{GeV}}\right]^{2}\,
       \frac{\left|\theta_{1}-\pi\right|}{\N}\, \mathcal{F}_{\rm kin, 1}(T_{1})\,,
       \label{Eq: energy density Kinetic+trapped only m1-1}
       \end{align}
       where $\mathcal{F}_{\rm kin, 1}(T_{1})$  is a function of the degrees of freedom at play, which can be found in Appendix~\ref{App: ratios gs grho} together with further details.  
       The relic density depends on the temperatures of all the different  SM copies via 
       the value of $m_1$ (i.e.~the  $\gamma_k$ values, see Eqs.~(\ref{gammalong}) and (\ref{ak1approx})). In the simplified case in which all the copies of the SM have the same temperature $T_{k \neq 0} \sim T^{\prime}$,  Eq.~(\ref{Eq: energy density Kinetic+trapped only m1-1}) can be written --for large $\N$ and for the largest possible kinetic misalignment contribution\footnote{Here the maximum axion velocity at the end of the trapped period is considered, i.e.~twice the mean axion velocity in \cref{Eq Mean theta dot}.}-- as
         \begin{align}
       \frac{\rho_{a,0}}{\rho_{\rm DM}}\bigg|_{\rm tr+kin,\, max} \simeq 0.64 
       \left(\frac{\,f_a}{ 10^{9}\, \mathrm{GeV}}\right)^{5/6}\,
       \frac{z^{\N/2}}{\N^{1/3}} \,\left|\theta_{1}-\pi\right|\,\mathcal{F}_{\rm kin,\,2}(T_{1}) \, , 
       \label{Eq: energy density Kinetic+trapped equal gamma Nfa}
       \end{align}
                where   $ \mathcal{F}_{\rm kin,\,2}(T_{1})$ is another function of the degrees of freedom at play, see Appendix~\ref{App: ratios gs grho}.

It also follows from the results above that  the trapped+kinetic DM solutions   behave in the $\{m_a,1/f_a\}$ plane as 
   \begin{align}
       \frac{\rho_{a,0}}{\rho_{\rm DM}}\bigg|_{\rm tr+kin} \propto m_a f_a^{19/12}\,,
       \label{mafatrki}
       \end{align}
   in contrast to  $m_af_a={\rm const.}$ for the pure trapped case in  Eq.~(\ref{Eq: pure trapped relic density}). The trapped+kinetic trajectories that account for DM are illustrated by  double lines in Fig.~\ref{fig:axionDM}, for 
   different values of the initial misalignment angle $\theta_1$.

\subsubsection{Solving both  the strong CP problem and the nature of DM}
\label{sec:CPconsminimum} 
We estimate here the minimum kinetic energy necessary in the trapped regime for the axion to roll over at least $\mathcal{O}(\N)$ maxima after $T_{\rm QCD}$. The result will be of general interest whenever $f_a<f_a^{\rm crit}$,  and of practical relevance mainly for the pure trapped misalignment. 

  Since the trapping forces $\theta_{\rm tr}\simeq \pi$,  the corresponding total energy density is always larger than the maximum potential energy (the height of the barrier $ 2 {m_{a}^2{f_a^2}}/{\N^2}$). As the axion rolls over $\sim \N$ different maxima, the energy density is diluted due to the expansion of the Universe,
  \begin{align}
  \rho_{a}(a_{{\rm QCD}+\N})= \rho_{a,{\rm tr}}\left(\frac{a_{\rm QCD}}{a_{{\rm QCD}+\N}}\right)^p\,,
  \end{align}
where $a_{{\rm QCD}+\N}$ is the scale factor  when the axion has rolled over just $\N$ maxima after $T_{\rm QCD}$, and $p$ parametrizes 
how strong is the dilution (we find numerically that $p \sim 1.5$). The minimal kinetic energy $K_{\rm min}$ necessary for the axion to be able 
 to roll over just $\N$ maxima before oscillating  must saturate the condition
\begin{align}
\left(K_{\rm min}+ 2 \frac{m_{a}^2{f_a^2}}{\N^2}\right)\left(\frac{a_{\rm QCD}}{a_{{\rm QCD}+\N}}\right)^p= 2 \,\frac{m_{a}^2{f_a^2}}{\N^2}\,.
\label{Eq kmin}
\end{align} 
Given the Hubble parameter in a radiation dominated Universe $H(t)=\frac{1}{2t}$  and the Friedmann equation in \cref{Eq:friedmann equation}, it  follows that the temperature $T_{{\rm QCD}+\N}$ after  overcoming $\N$ maxima (corresponding to a time difference of $\delta t\sim 2\pi\N/m_a$, see discussion before Eq.~(\ref{closest})) is 
 \begin{align}
 T_{{\rm QCD}+\N}=\left[\frac{4\pi^3}{45}g_*(T_{\rm QCD})\right]^{1/4}\sqrt{\frac{M_{\rm Pl}}{2(t_{\rm QCD}+{2\pi\N}/{ m_a})}} 
 \simeq T_{\rm QCD}- 8.2\,\frac{2\pi\N}{ m_a}\frac{T_{\rm QCD}^3}{  M_{\rm Pl}}\,,
 \end{align}
  where $t_{\rm QCD}$ is the time corresponding to $T_{\rm QCD}$, and the approximation  ${2\pi\N}/{ m_a}\ll t_{\rm QCD}$\footnote{Or equivalently ${2\pi\N\, T^2_{\rm QCD} } \ll {m_a\,M_{\rm Pl}}$.} has been used, as in this regime $H\ll m_a$.
 The corresponding dilution factor reads 
  \begin{align}
 \left(\frac{a_{\rm QCD}}{a_{{\rm QCD}+\N}}\right)^p
 \simeq\left(\frac{T_{{\rm QCD}+\N}}{ T_{\rm QCD}}\right)^p
 \simeq 1-8.2\,p\,\frac{2\pi\N}{ m_a}\frac{T_{\rm QCD}^2}{  M_{\rm Pl}}\,.
 \end{align}
 Substituting it in \cref{Eq kmin}, the minimum kinetic energy required is determined, 
 \begin{align}
 K_{\rm min}\simeq 8.2\,p \,\frac{4\pi}{\N}\,\frac{ m_a  T_{\rm QCD}^2}{ M_{\rm Pl}} f_a^2\,.
 \end{align}
To evaluate how probable it is to simultaneously solve the strong CP problem and explain DM,  $ K_{\rm min}$ needs to be compared with the mean kinetic energy generically expected at  $T_{\rm QCD}$. It follows from Eq.~(\ref{Eq Mean theta dot}) that
     \begin{align}
   K_{\rm tr}\equiv   \frac{1}{2}{\langle \dot \theta^2_{\rm tr}\rangle}\,f_a^2=   \frac{1}{4} {m_1 m_{a,\pi}^{\rm QCD}} \left(\frac{a_1}{a_{\rm QCD}}\right)^{3} (\theta_{1}-\pi)^2 f_a^2\,.
      \label{Eq Mean theta dot2}
      \end{align} 
In the case in which all mirror copies but the SM have the same temperature $T'$ 
and taking for example $\gamma'=0.2$, the latter equation can be rewritten as 
 \begin{align}
 K_{\rm tr}\simeq \frac{1}{4} 
 \left(\frac{T_{\rm QCD}^{14}(m_{a,\pi}^{\rm QCD})^5 }{M_{\rm Pl}^7}\right)^{1/6}\,(\theta_{1}-\pi)^2\,f_a^2\times \kappa\,, 
 \end{align}
 where $\kappa$ is a numerical factor that ranges in the interval $(4-8)$. This implies that the axion field will roll over at least $ \N$ barrier tops whenever 
   \begin{align}
  m_a\lesssim \frac{\N}{4\pi \, 8.2 \, p}\left(\frac{T_{\rm QCD}}{\sqrt{M_{\rm Pl}}}\right)^{1/3}(m_{a,\pi}^{\rm QCD})^{5/6}|\theta_1-\pi|^2\,. 
  \label{Eq: condition roll over N maxima}
  \end{align}
    We compare next the parametric dependence of all DM axion solutions --which also solve the strong CP problem-- discussed up to this point. 
They are depicted in the $\{m_a,1/f_a\}$ plane of Fig.~\ref{fig:axionDM} and for several values of 
$\theta_1$:  
  
\begin{figure}[ht]
\centering
\includegraphics[width=0.9\textwidth]{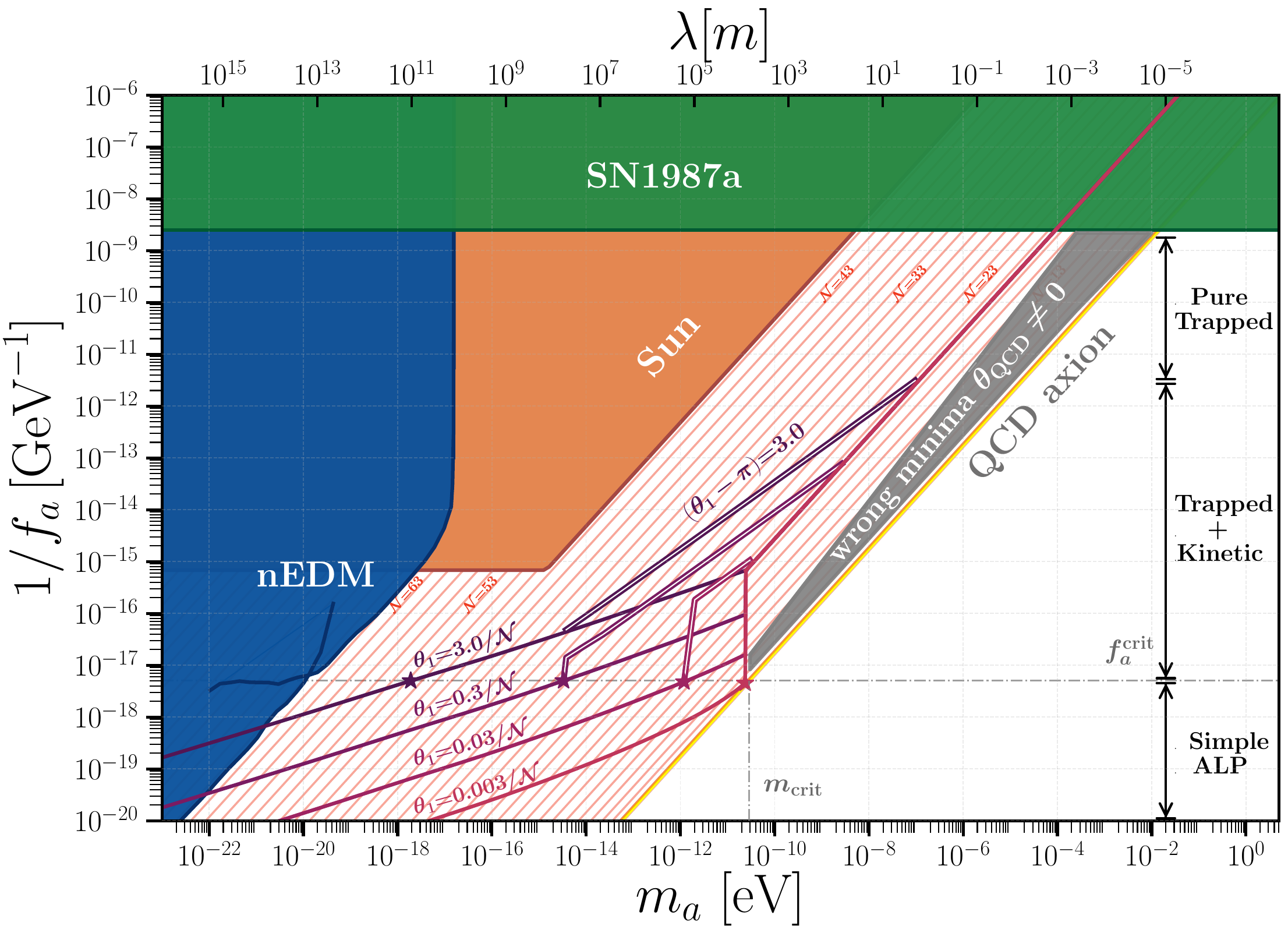}
\caption{ $Z_\N$ axion solutions that solve both the strong CP problem and account for DM. The superimposed single continuous lines indicate either  simple ALP solutions or pure trapped ones, while double lines denote trapped+kinetic misalignment regions.  
The tenuous oblique lines for various values of $\N$  indicate $Z_\N$ axion solutions to the strong CP problem. 
Solid colors denote regions experimentally excluded by axion-nEDM~\cite{Graham:2013gfa} or astrophysical data. Further constraints and prospects can be seen in Fig.~\ref{fig:axionEDM}.}  
\label{fig:axionDM}       
\end{figure}

  \begin{itemize}
 \item  In the excluded  grey area the strong CP problem is not solved since the condition in \cref{Eq: condition roll over N maxima} is not satisfied. This precludes any DM solution with $\N\lesssim 9$, although only for $m_a\gtrsim m_{\rm crit}\simeq 4\times 10^{-11}\, \text{eV}$.
 \item The stars indicate for each trajectory the point below which $f_a$ is large enough ($f_a>f_a^{\rm {crit}}$) so that the oscillations  around the true minimum  start after the zero temperature potential is fully developed. This is the  
``simple ALP''
case with DM solutions obeying $m_af_a^4 \propto  {\rm const.}$
 \item Above the stars,  double (single) continuous lines indicate trapped+kinetic (pure trapped) trajectories.   For large values of $f_a$ within the $f_a<f_a^{\rm {crit}}$ range, the trapped+kinetic solution tends to dominate the parameter space, while for small $f_a$ the pure trapped mechanism sets the relic density. 
 \item The kinetic kick received by a given trapped+kinetic trajectory when its $f_a$ value  is smaller than $f_a^{\rm crit}$ (but very close to it)  
  can be clearly seen. Above that point, the slope of the trajectory obeys $m_a f_a^{19/12} ={\rm const.}$, as explained around \eq{mafatrki}.
 \item The pure trapped regime, that populates the top of the figure for small enough $f_a$, 
 corresponds to the continuous $\N\sim 21$ line, whose slope obeys  $m_a f_a ={\rm const.}$
  \end{itemize}
  Overall, Fig.~\ref{fig:axionDM} shows that good solutions to both the strong CP problem and the nature of DM exist within the $Z_\N$ axion scenario whenever $m_a\lesssim 10^{-4}$ eV,  down to the fuzzy DM region ($m_a \sim 10^{-22}$ eV). In the range   $ m_{\rm crit}\lesssim m_a \lesssim 10^{-4}$ eV the number of worlds must be $ \N\sim21$.  For the lighter axion scenarios,  $m_a \lesssim m_{\rm crit}$, values of $\N$ up to $\sim65$ are possible.   

Finally, note that this analysis has focused only on the axion zero mode as a first step. Within the trapped mechanism the axion field  self-interactions may become important whenever the axion is not close to the minimum. Non-linearities will induce the production of higher momentum axion quanta, i.e.~axion fragmentation \cite{Fonseca:2019ypl}. The consequences of the fragmentation of the axion field within the trapped misalignment mechanism is left for a future work.

\subsection{Trapped vs.~non-trapped mechanisms} 
\label{subsec:parametrics_of_the_differnt_production_mechanisms}

The trapped mechanism identified in this paper is actually 
more general than the axion setup 
with a non-linearly realized $Z_\N$ shift symmetry discussed above. 
Trapped misalignment could arise in fact in a large variety of QCD axion or ALP scenarios.

  In QCD axion models, any additional PQ-breaking source 
  at low energies 
  needs to be extremely suppressed in order to comply with the nEDM bounds. However, the 
  high-temperature axion
  potential is largely unconstrained. Indeed, any PQ-breaking  source which is  active only at high energies may induce a displacement of the potential minima at high temperatures.  The axion field can then get trapped in its displaced minimum for a certain period of time, until the low-energy QCD potential develops. This can dramatically change the relic density predicted today, due to the trapped misalignment phase. An enhancement can be expected whenever the transition between the two phases of the potential is non-adiabatic\footnote{In regimes which are mainly adiabatic, 
  a reduction of the axion relic density may result instead~\cite{Nakagawa:2020zjr}.} and takes place at a smaller temperature than the oscillation temperature in absence of trapping.
  \begin{table}
    \begin{align}
      \vline \begin{array}{l|ll} \hline\\[-7pt] \centering
      \text{Misalignment mechanisms}\quad&\qquad \rho_{a,0} \quad&\quad  g_{aXX}\propto {1}/f_a \\[5pt]
        \hline \\[-10pt]
      \text{Simple ALP}     \quad&\quad   \sqrt{m_a}f_a^2  \quad&\qquad {\centering m_a^{1/4}}\\[10pt]
      \text{Trapped} \quad&\quad   m_a^2f_a^2  \quad&\qquad m_a\\[8pt]
      \text{Trapped+Kinetic} \quad
              &\quad   m_af_a^{19/12}  \quad
              &\qquad m_a^{12/19}\\[5pt]
      \hline\hline \\[-10pt]
      \text{QCD-like }   \quad&\quad   m_a^{5/6}f_a^{2}  \quad&\qquad m_a^{5/12}\\[5pt]
      \hline
      \end{array}\vline
    \end{align}
      \caption{Parametric dependence of axion relic density and axion couplings 
      $g_{aXX}$ (for a fixed axionic relic density), for the different types of misalignment 
      mechanisms discussed in the text. 
      }
      \label{tablemafa}
  \end{table}
  The parametric dependence of the axion relic density on the $\{m_a, f_a\}$ plane is a hallmark of the different misalignment mechanisms discussed above. Table~\ref{tablemafa} summarizes the results presented earlier on (see Eqs.~(\ref{mafaALP}), (\ref{Eq: pure trapped relic density}) and (\ref{mafatrki})). It also shows the generic $m_a$ dependence of the axion-SM interaction couplings $g_{aXX}\sim 1/f_a$ (with $X$ denoting a generic SM field) 
  for each DM production scenario discussed above. 
  For comparison, the corresponding results for the canonical QCD axion scenario are shown as well (within the $Z_\N$ axion there is no regime where the QCD-like relic density formula applies other than for $\N=1$). 
 In Table~\ref{tablemafa} (and in all figures), 
 for the QCD-like and trapped+kinetic cases the DIGA value $\alpha=8$ has been assumed 
 (see \cref{h}).\footnote{
  For arbitrary $\alpha$, $\rho_a\propto m_af_a^{{(14+3\alpha)}/{(2(4+\alpha))}}$ in the trapped+kinetic case, while 
  $\rho_a\propto m_a^{(\alpha+2)/(\alpha+4)}f_a^2$  in QCD-like misalignment, from which the corresponding $m_a$ dependence of $g_{aXX}\propto {1}/f_a$  can be obtained.
}

\begin{figure}[ht]
\centering
\includegraphics[width=0.75\textwidth]{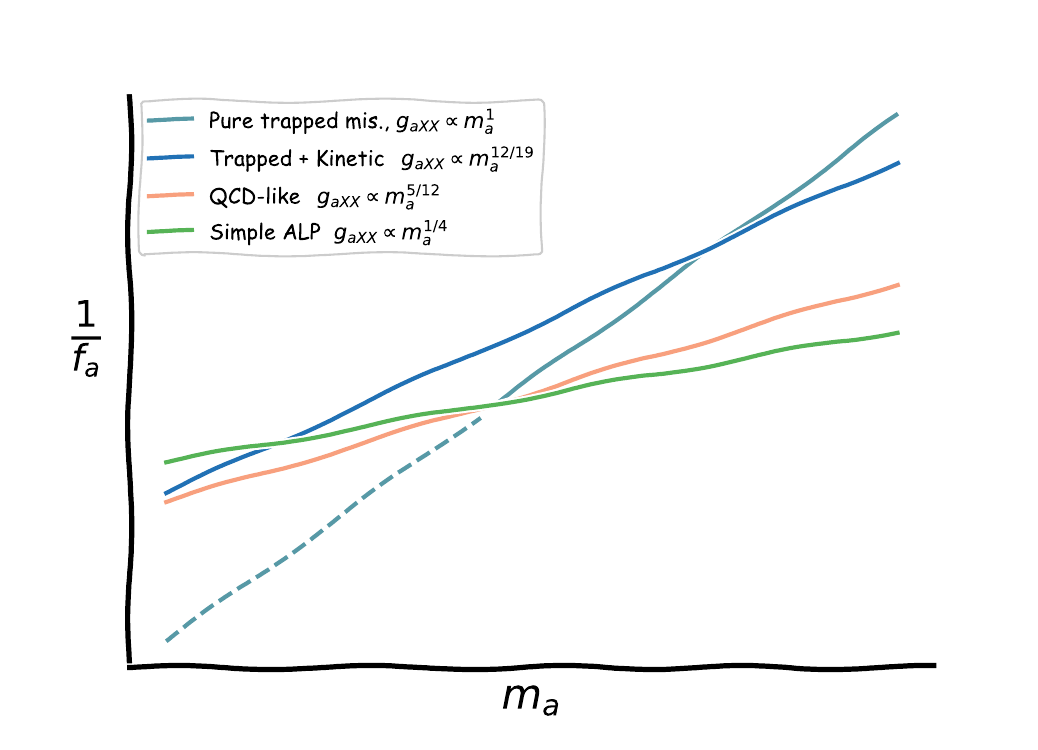} 
\caption{Sketch of the parametric dependence of $g_{aXX}(m_a)$ for a fixed axionic relic density, for the different types of misalignment discussed. For the QCD-like and the pure trapped cases, the DIGA-inspired power-law temperature suppression is assumed, i.e.~$\alpha=8$. The dashed line  reminds that the pure trapped  $Z_\N$ scenario can only take place for $m_a>m_{\rm crit}$. 
}
\label{fig:Sketch}       
\end{figure}

The differences in slope shown in  Table~\ref{tablemafa} 
are illustrated qualitatively in Fig.~\ref{fig:Sketch}. The crossing point of all lines but the trapped+kinetic misalignment trajectory corresponds to $m_{\rm crit}$.  The figure shows intuitively how the experimental parameter space is populated differently in the trapped  mechanisms (blue tones)
 compared to that of the QCD axion ({\color{orangeQCD} \bf orange}). For a given $m_a$ to which an experiment is sensitive, the value of the signal strength expected  $g_{aXX}\sim1/f_a$ is generically larger in the trapped regimes, and thus the experimental reach is higher.  

\section{Implications for axion dark matter searches}
\label{sec:pheno}

In this section we discuss the implications of the above $Z_\N$  scenario   
for axion DM searches, namely those experiments that rely on the 
hypothesis that the axion sizeably contributes to the DM relic density. 
More specifically,  the $Z_\N$ axion coupling to photons, 
nucleons, electrons and to the nEDM will be discussed. 
The whole mass range from the canonical QCD axion mass down to the ultra-light QCD axion regime (with masses $m_a \ll 10^{-10}$ eV) is considered, including the fuzzy DM regime 
down to $m_a\sim10^{-22}$ eV. 

All the couplings considered but the axion-nEDM coupling 
are model dependent: they can be enhanced or suppressed in specific UV completions of the axion paradigm. We will  explore for each of those    
 the parameter space coupling vs.~$m_a$. On the right-hand side of the figures, though, the naive expectations for $f_a$ when assuming 
 $\mathcal{O}(1)$  couplings will be indicated as well. This is done for pure illustrative purposes as the relation between $f_a$ and axion couplings is model-dependent. 
 
  In contrast, the value of the axion-nEDM coupling  only depends on $f_a$, that is, it only assumes that the axion solves the strong CP problem. 
   The same model independence holds for previous analyses of highly-dense stellar objects and gravitational wave prospects, which in addition did not need to assume an axionic nature of DM. These efforts  led to strong constraints on the $\{m_a, f_a\}$ parameter space~\cite{Hook:2017psm,Huang:2018pbu,ZNCPpaper}, 
   which can be thus directly compared with the prospects for axion-nEDM searches.
   
The figures below depict with solid (translucent) colors the experimentally excluded areas of parameter space (projected sensitivities).
{\it The blue
tones are reserved exclusively for experiments which do rely on the assumption that axions  account sizeably for DM,} while the remaining colors indicate searches which are independent of the nature of DM.  In case the axion density  $\rho_a$ provides only a fraction of the total DM relic density $ \rho_{\rm DM}$, 
the sensitivity to couplings of axion DM experiments
should be rescaled 
by 
$(\rho_a / \rho_{\rm DM})^{1/2}$.

Crucially, we have shown in \sect{sec:axion_dark_matter}  
that in some regions of the $\{m_a, f_a\}$ plane 
the $Z_\N$ axion can realize 
DM via the misalignment mechanism 
and variants thereof,  
depending on the possible cosmological histories of the 
axion field evolution in the early Universe. The identified regions will be superimposed on the areas of parameter space experimentally constrained/projected by axion DM experiments.

In order to compare the experimental panorama with the predictions of a benchmark axion model, the figures  will  also show the expectations for the coupling values within the $Z_\N$-KSVZ axion model
 developed in Ref.~\cite{ZNCPpaper}. 
 The canonical KSVZ QCD axion solution (i.e.~$\N=1$) is shown as 
 a thick yellow line, 
 embedded into a faded band encompassing 
 the model dependency of the KSVZ axion \cite{DiLuzio:2016sbl,DiLuzio:2017pfr}.
 Oblique orange lines will signal instead the center of the displaced yellow band that corresponds to solving the strong CP problem with other values of $\N$, that is,  for a $Z_\N$ reduced-mass axion, see Eq.~(\ref{maZNLargeN}).  

The entire DM relic density can be accounted for  within the $Z_\N$ axion paradigm in the regions  
encompassed by the purple band in the figures.\footnote{The band's width accounts qualitatively for corrections to the analytic solutions  obtained in the previous section, see e.g.~comment after Eq.~(\ref{N21}) and Fig.~\ref{fig:axionDM}. For instance, in the trapped regime a factor of 2 uncertainty on $f_a$ has been applied. } 
These correspond 
to  initial values of $\theta_a$ (from the misalignment mechanism) which range from 
$\theta_1= 3/\N$  down to $\theta_1=0.003/\N$.  
 The figures illustrate that in the pure trapped regime the relic density is independent of the initial misalignment angle. In contrast, for  the simple ALP and the trapped+kinetic mechanisms it does depend on the value of $\theta_1$ (in the latter case, through its dependence on the axion velocity at $T_{\rm QCD}$).

\subsection{Axion coupling to photons}
\label{sec:axionphoton}

\begin{figure}[!th]
\centering
\includegraphics[width=0.9\textwidth]{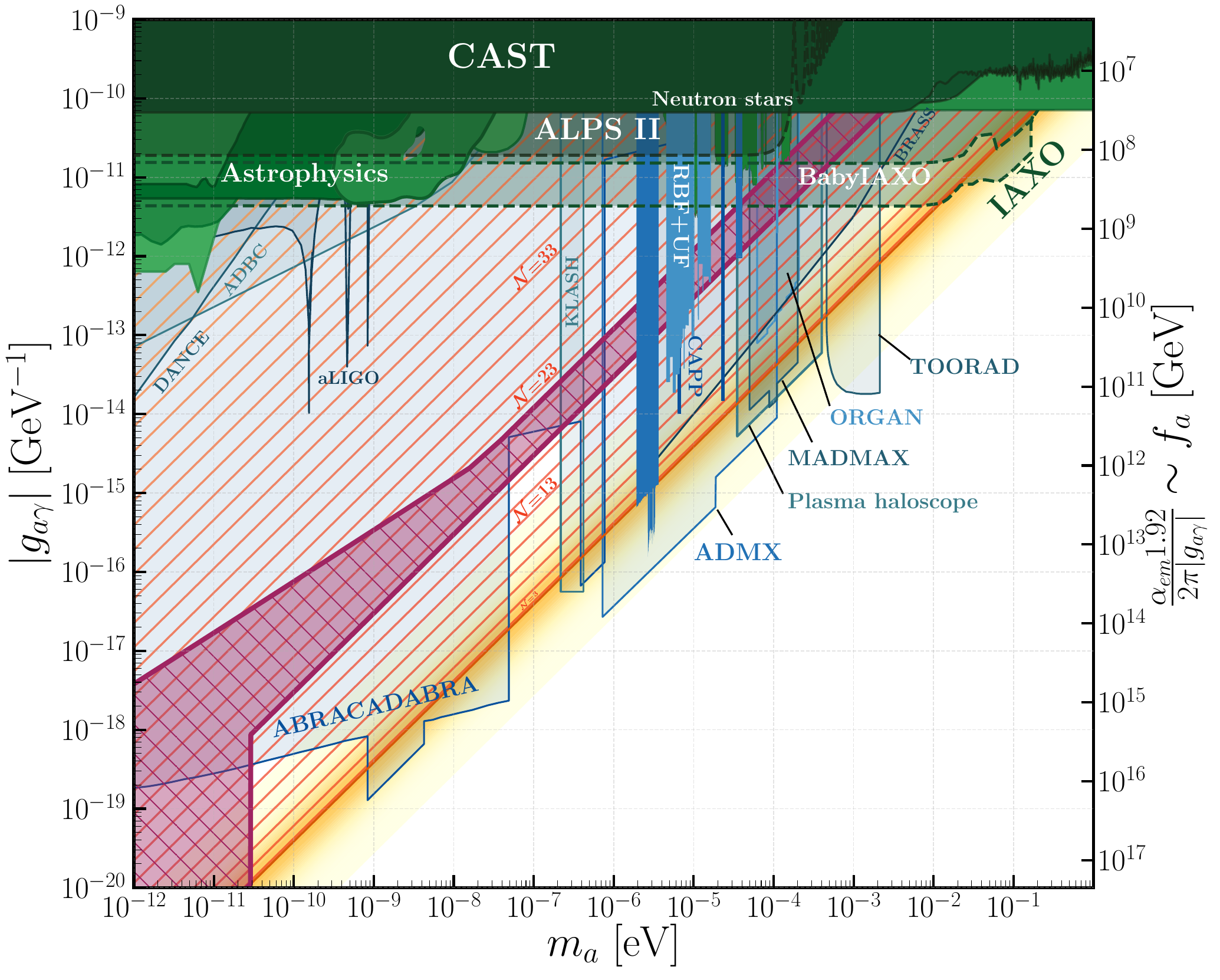} 
\caption{Axion-photon coupling vs.~axion mass. 
Axion limits readapted from \cite{ciaran_o_hare_2020_3932430} include: 
laboratory axion experiments and helioscopes ({\color{colLab} \bf dark green}) \cite{Bahre:2013ywa,Armengaud:2014gea,Anastassopoulos:2017ftl,Abeln:2020ywv}, axion DM experiments ({\color{colnEDM} \bf blue}) \cite{Hagmann:1990tj,McAllister:2017lkb,Alesini:2017ifp,Zhong:2018rsr,Marsh:2018dlj,Braine:2019fqb,Lee:2020cfj,Beurthey:2020yuq,Alesini:2020vny,Lawson:2019brd} and astrophysical bounds ({\color{colAstro} \bf green})~\cite{Abramowski:2013oea,Ayala:2014pea,Payez:2014xsa,TheFermi-LAT:2016zue,Marsh:2017yvc}. 
Projected sensitivities appear in translucent colors.  
The {\color{colZN} \bf orange} oblique lines represent the theoretical prediction for solutions to the strong CP problem within the $Z_\N$ paradigm, for the benchmark axion-photon couplings 
$E/N=0$ and for different (odd) numbers of SM copies $\N$. 
In  the {\color{purpleDMbands} \bf purple} band area 
the $Z_\N$ axion can account for the entire DM density, in addition to solve the strong CP problem. 
}
\label{fig:axionphoton}       
\end{figure}

The effective axion-photon-photon coupling $g_{a\gamma}$ is defined via the Lagrangian 
\begin{equation}
\delta \mathcal{L} \equiv \frac{1}{4} g_{a\gamma} a F \tilde F\,,
\end{equation}
 with \cite{diCortona:2015ldu} 
\beq\label{agammagamma_coupling}
g_{a\gamma} = \frac{\alpha}{2\pi f_a} (E/N - 1.92(4)) \, ,
\eeq
where $E/N$ takes model-dependent values ($E$ and $N$ denote the model-dependent anomalous electromagnetic and strong contributions). This coupling 
is being explored by a plethora of experiments, as illustrated in \fig{fig:axionphoton}. 
For reference, 
predictions of the benchmark $Z_\N$-KSVZ axion model~\cite{ZNCPpaper} are depicted for $E/N=0$.
The conclusion is that, for a large range of values of $\N$, an axion-photon signal can be expected in large portions of the parameter space of upcoming axion DM experiments, if the $Z_\N$ axion accounts for the entire DM relic density.

\subsection{Axion coupling to nucleons}
\label{sec:axionnucleon}

\begin{figure}[!th]
\centering
\includegraphics[trim=0 0 0 145,clip,width=0.9\textwidth]{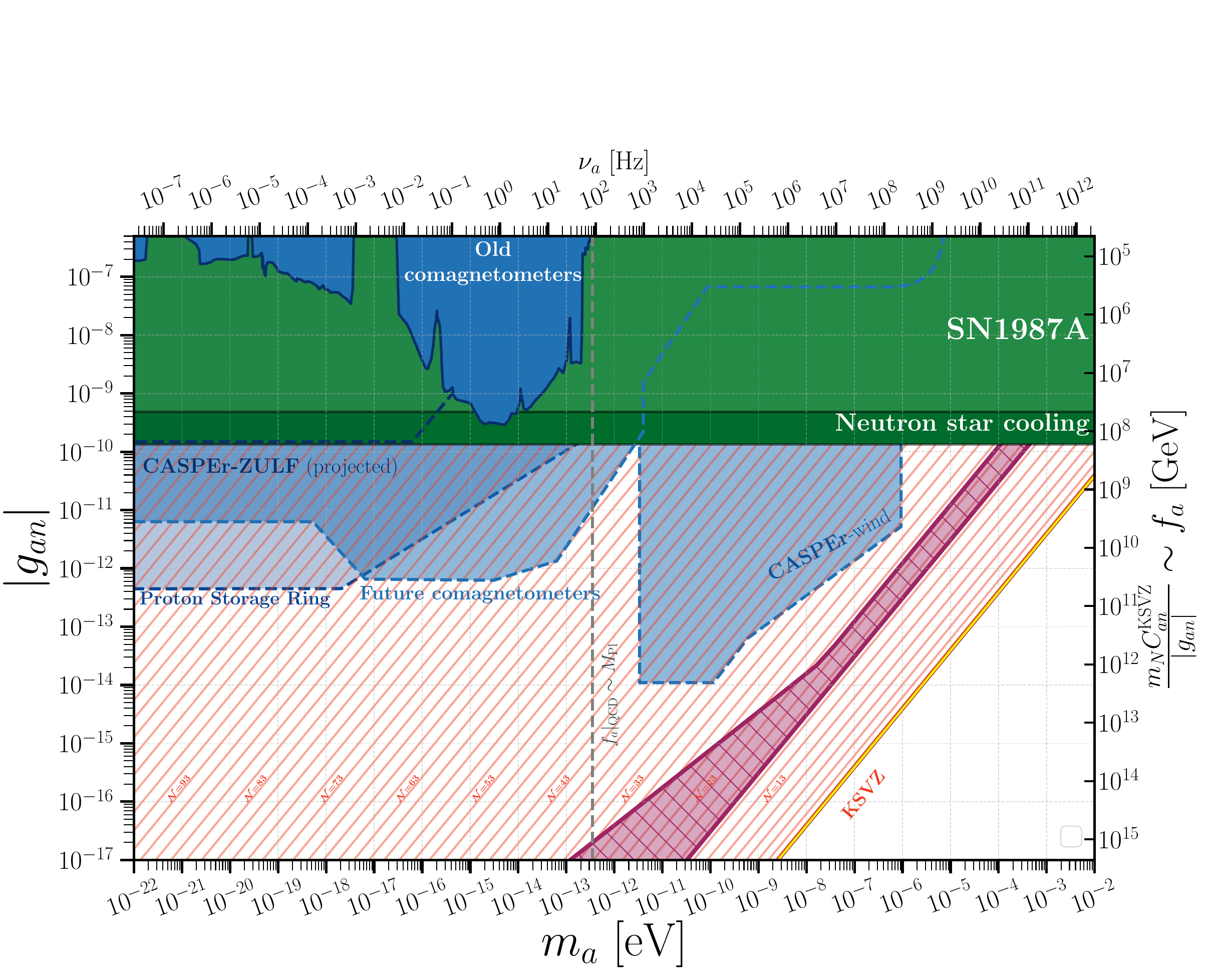}
\caption{Axion-neutron coupling vs.~axion mass. 
Axion limits readapted from \cite{ciaran_o_hare_2020_3932430} include: 
axion DM experiments ({\color{colnEDM} \bf blue}) \cite{JacksonKimball:2017elr,Wu:2019exd,Garcon:2019inh,Bloch:2019lcy,Graham:2020kai} and astrophysical bounds ({\color{colAstro} \bf green}) \cite{Beznogov:2018fda,Carenza:2019pxu}. 
Projected sensitivities appear in translucent colors,   
delimited by dashed lines.  
The  {\color{colZN} \bf orange} oblique lines represent the theoretical prediction for the $Z_\N$ axion-neutron couplings,
for different (odd) numbers of SM copies $\N$. 
The {\color{purpleDMbands}\bf purple} band encompasses the region where the $Z_\N$ axion  can account for the entire DM density. The two theoretical predictions are represented for the benchmark value $c^0_q = 0$.} 
\label{fig:axionneutron}       
\end{figure}

The axion coupling to nucleons $N$ ($N=p,n$),  
\beq
g_{aN} = C_{aN} m_N / f_a\,,   
\label{aN_coupling}
\eeq 
is defined via the Lagrangian term 
\beq 
\delta \mathcal{L} \equiv C_{aN} \frac{\partial_\mu a}{2 f_a} 
\bar N \gamma^\mu \gamma_5 N\,,  
\eeq
with~\cite{diCortona:2015ldu} (see also \cite{Vonk:2020zfh} for similar results)
\begin{align}
\label{eq:Cap}
C_{ap} &= -0.47(3) + 0.88(3) \, c^0_u - 0.39(2) \, c^0_d - C_{a,{sea}}
\, , \\
\label{eq:Can}
C_{an} &= -0.02(3) + 0.88(3) \, c^0_d - 0.39(2) \, c^0_u - C_{a,{sea}} \,,
\end{align}
where $C_{a,{sea}} = 0.038(5) \, c^0_s 
+0.012(5) \, c^0_c + 0.009(2) \, c^0_b + 0.0035(4) \, c^0_t$ denotes a ``sea-quark'' 
contribution in the nucleon, while in the reference $Z_\N$-KSVZ axion model 
$c^0_q = 0$ for all quark flavours $q$.  
The parameter space of the axion-nucleon coupling vs.~axion mass 
is displayed in \fig{fig:axionneutron}. It demonstrates that 
no signals can be expected in the axion DM experiments foreseen up to now.

\subsection{Axion coupling to electrons}
\label{sec:axionelectron}
\begin{figure}[ht]
\centering
\includegraphics[width=0.9\textwidth]{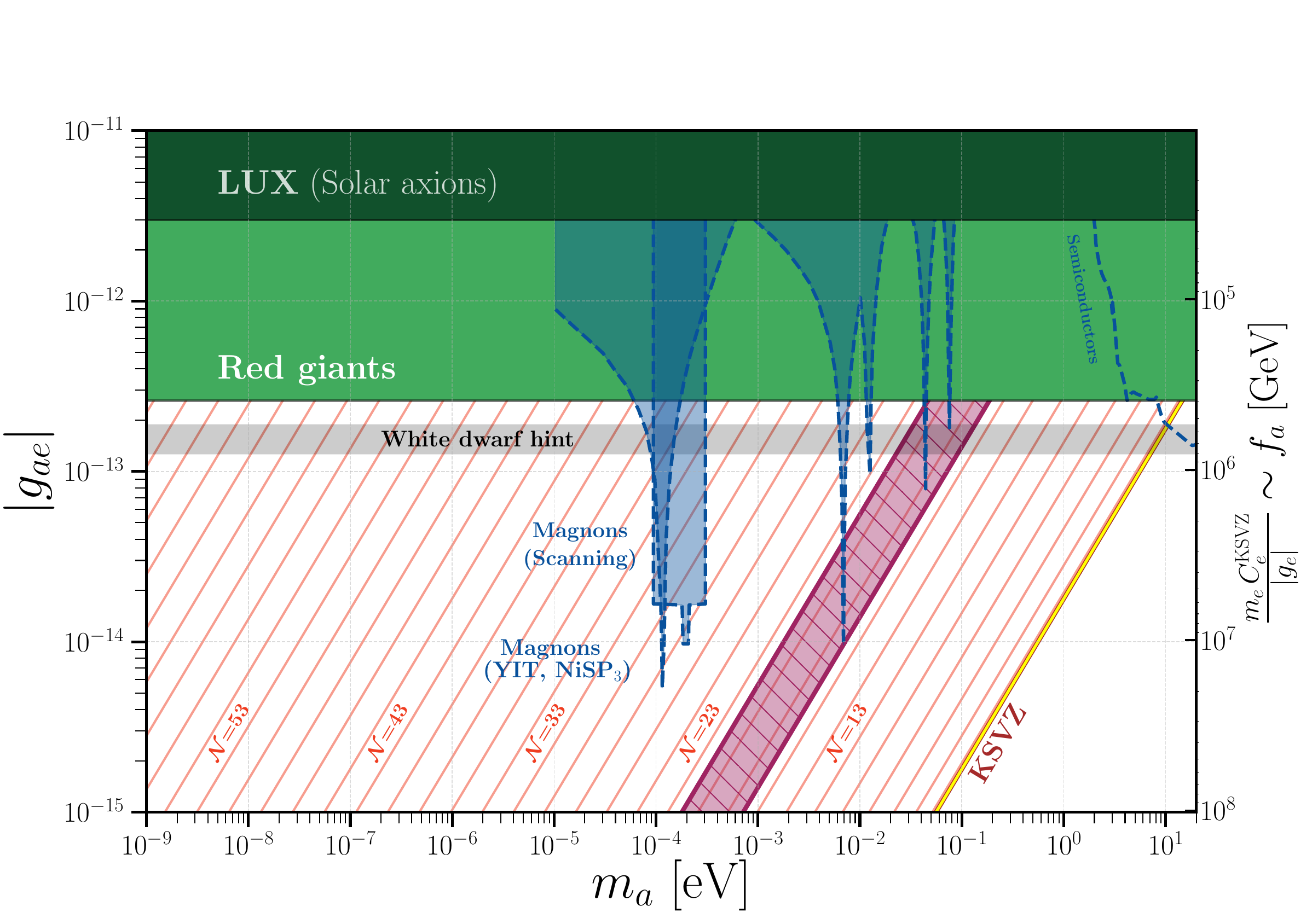} 
\caption{Axion-electron coupling vs.~axion mass. 
Axion limits readapted from \cite{ciaran_o_hare_2020_3932430} include: 
solar axion experiments ({\color{colLab} \bf dark green})~\cite{Bloch:2016sjj,Akerib:2017uem}, 
axion DM experiments ({\color{colnEDM} \bf blue}) \cite{Chigusa:2020gfs,Mitridate:2020kly} and astrophysical bounds 
({\color{colAstroL} \bf green}) \cite{Capozzi:2020cbu} and hints (grey band) \cite{Giannotti:2017hny}. 
Projected sensitivities appear in translucent colors,   
delimited by dashed lines.  
The {\color{colZN} \bf orange} oblique lines represent the theoretical prediction for the $Z_\N$ axion-electron couplings assuming $c^0_e = 0$ 
and $E/N=0$, for different (odd) numbers of SM copies $\N$. 
The {\color{purpleDMbands}\bf purple} band encompasses the region where the $Z_\N$ axion can account for the entire DM density.
}
\label{fig:axionelectron}       
\end{figure}

The axion coupling to electrons, 
\beq
g_{ae} = C_{ae} m_e / f_a\,,
\label{ae_coupling}
\eeq 
is defined via the Lagrangian term 
\beq
\delta \mathcal{L} \equiv C_{ae} \,\frac{\partial_\mu a}{2 f_a} \, \bar e \gamma^\mu \gamma_5 e\,, 
\eeq
with \cite{Srednicki:1985xd,Chang:1993gm}
\beq 
C_{ae} = c_e^0 + \frac{3\alpha^2}{4\pi^2} 
\[ \frac{E}{N} \log\( \frac{f_a}{m_e} \)
- 1.92(4)
\log\( \frac{ {\rm GeV}}{m_e} \) \] \, ,
\label{Cae_KSVZ_ZN} 
\eeq
where in the reference $Z_\N$-KSVZ axion model 
$c^0_e = 0$ and $E/N = 0$. 
The parameter space of the axion-electron coupling vs.~axion mass 
is displayed in \fig{fig:axionelectron}. It shows that axion-magnon 
conversion techniques~\cite{Chigusa:2020gfs,Mitridate:2020kly} 
could barely detect a $Z_\N$ reduced-mass axion which would account for DM. It would be interesting to explore whether  experiments which probe the axion electron coupling with different techniques, such as QUAX \cite{Crescini:2020cvl}, will be able to test this scenario in the future.

\subsection{Axion coupling to nEDM}
\label{sec:axionnEDM}

The axion coupling to the nEDM operator provides a crucial test of the 
$Z_\N$ axion setup: it only depends on $f_a$,  
 that is, on the assumption that the axion solves the strong CP problem, see Eq.~(\ref{LnEDM}). Unlike the couplings discussed above, it {\it does not} depend on the details of the UV completion of the axion model. The signal thus depends solely on $f_a$ and $m_a$. 

In fact, standard mechanisms to enhance 
QCD axion couplings via large Wilson coefficients do not modify the size of the  nEDM coupling, which is fixed purely by the QCD $m_a$-$f_a$ relation. 
In contrast, the $Z_\N$ axion setup under discussion 
is responsible for a ``universal'' enhancement of all axion couplings 
(including that to the nEDM operator), for any given $m_a$ mass, 
following the rescaling with $\N$ 
of the modified $m_a$-$f_a$ relation in \eq{maZNLargeN}.

\begin{figure}[ht]
\centering
\includegraphics[width=0.9\textwidth]{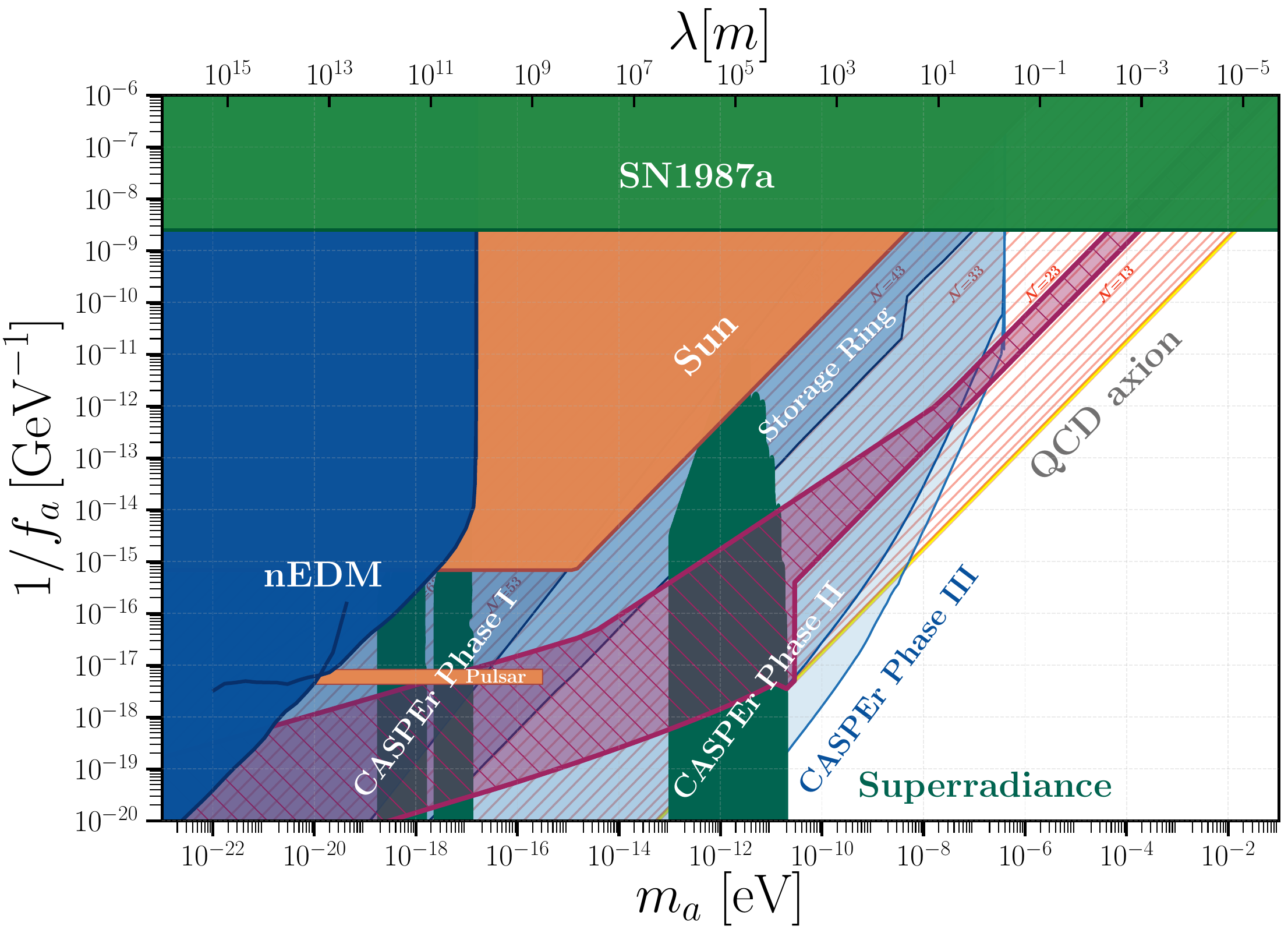}
\caption{Model-independent constraints in the $\{m_a, f_a\}$ plane, 
including axion DM experiments ({\color{colnEDM} \bf blue}) \cite{Graham:2013gfa,Chang:2017ruk,JacksonKimball:2017elr} 
astrophysical ({\color{colAstro} \bf green}) \cite{Raffelt:2006cw,Arvanitaki:2014wva,Arvanitaki:2010sy,Stott:2020gjj,Mehta:2020kwu} and finite-density constraints ({\color{colFinDen} \bf orange}) \cite{Hook:2017psm,Huang:2018pbu,ZNCPpaper}.  The nEDM bounds have been relaxed by a factor of 3 to take into account the stochastic nature of the axion field, see Ref.~\cite{Centers:2019dyn}. 
Projected sensitivities appear in translucent colors (those relative to finite-density effects can be 
seen in Fig.~8 of \cite{ZNCPpaper}). 
The {\color{purpleDMbands}\bf purple} band encompasses the region where the $Z_\N$ axion can account for the entire DM density. }
\label{fig:axionEDM}       
\end{figure}

Were the relic DM density to be entirely
 made up of axions, 
 an oscillating nEDM signal $d_n (t)$ could be at reach \cite{Graham:2013gfa},
\beq 
d_n (t) = g_{d}\,\frac{\sqrt{2\rho_{\text{DM, local}}}}{m_a} \cos (m_a\,t) \, ,
\eeq
where 
$g_{d} = C_{an\gamma} / (m_n f_a)$
is the coupling of the axion to the nEDM operator, 
defined via the Lagrangian term 
\beq
\label{LnEDM}
\delta \mathcal{L} \equiv - \frac{i}{2} \frac{C_{an\gamma}}{m_n} \frac{a}{f_a}  \bar n \sigma_{\mu\nu} \gamma_5 n F^{\mu\nu}\,,
\eeq
 where 
$C_{an\gamma} = 0.011(5) \, e$ \cite{Pospelov:1999mv} and  
$\rho_{\rm DM, \, local} \approx 0.4\ \text{GeV}/\text{cm}^3$ is the local energy density of 
axion DM.\footnote{Note that the local relic density differs from the \emph{mean} relic density as measured through the CMB and used in \cref{Eq:simple ALP relic dansity ratio CMB}.}  
For instance, the axion DM experiment 
CASPEr-Electric \cite{Budker:2013hfa,JacksonKimball:2017elr,Aybas:2021nvn} 
employs nuclear magnetic resonance techniques to search 
for an oscillating nEDM. 
Its reach in the $\{m_a, f_a\}$ parameter space 
is displayed in \fig{fig:axionEDM}.
The reach of other techniques 
to probe an oscillating nEDM based on 
storage rings \cite{Chang:2017ruk} are also illustrated there, together with present constraints which exclude $\N\gtrsim65$. 

\fig{fig:axionEDM} demonstrates the brilliant prospects of CASPEr-Electric Phase I and II to discover a  $Z_\N$ axion in a large range of light masses and for $3\le\N\lesssim65$ if such axion accounts for DM.
In particular, the 
$Z_\N$ axion paradigm with reduced mass is --to our knowledge-- the first axion model that could explain a positive signal in CASPEr-Electric Phase I (and large regions of Phase II), and at the same time solve the strong strong CP problem. Canonical QCD axion models that exhibit large enough nEDM couplings to be detectable at these experiments predict automatically too heavy axions and therefore out of their reach. 
 The figure  illustrates in addition that future proton storage ring facilities may access the region of interest, albeit only in a small parameter region with large $\N$ and $\theta_1$ values. 

Furthermore, data on highly dense stellar objects share with the oscillating nEDM signal its model-independence character: they directly provide constraints on the $\{m_a, f_a\}$ parameter space. Those data have the added value of not relying on an axionic nature for DM.  
They have already set strong constraints on the $\{m_a, f_a\}$ parameter space ~\cite{Hook:2017psm,Huang:2018pbu,ZNCPpaper}.  For the $Z_\N$  axion scenario under discussion,  they implied a strict  limit on the number of mirror worlds:  $3\le\N\lesssim47$ for $f_a\lesssim 2.4\times 10^{15}$ GeV. They also provide   tantalizing
  discovery prospects  for any value of $f_a$ and down to $\N\sim 9$ --possibly even lower--
  with future neutron star and gravitational wave data, down to the ultra-light mass region, see Fig.~8  in Ref.~\cite{ZNCPpaper} (BBN constraints \cite{Hook:2017psm} are not shown, as they do not overlap with the purple region where DM is accounted for).   It is thus of crucial interest to compare them with the oscillating nEDM projects described above. \fig{fig:axionEDM} tends to this task and shows that present finite-density data leave wide open the complete parameter space that CASPEr-Electric I, II and III can cover, with exciting synergy prospects between both type of approaches.

\subsection{Ultra-light QCD axion dark matter}
\label{sec:ULQCDA}

The very light 
pseudoscalar
DM candidates usually referred to as ultra-light axions, with mass in the range $m_a \in \[ 10^{-22}, 10^{-10}\]$ eV, 
do not customarily address the strong CP problem. This is because, for the canonical QCD axion, such light masses would imply trans-Planckian decay constants, $f_a>M_{\rm Pl}$.   In contrast, the $Z_\N$ axion scenario is --to our knowledge-- \emph{the first axion model of fuzzy DM that can also solve the strong CP problem}.

Ultra-light axions are typically searched via
gravitational interactions, with no reference to their 
highly suppressed couplings to the SM (for a recent White Paper see \cite{Grin:2019mub}). 
The $Z_\N$ scenario offers instead an interesting complementarity between strong CP related experiments and purely gravitational probes of fuzzy dark matter. 

Whenever the ultra-light DM  axions obey cosinus-like potentials, they are also subject to CMB  
constraints on the matter-power spectrum, due to the presence of anharmonic terms 
which impact the growth of perturbations (see e.g.~Refs.~\cite{Desjacques:2017fmf,Poulin:2018dzj}).
This implies a bound around the recombination era on any additional 
energy density fluctuations beyond cold DM, $\delta\rho / \rho \lesssim 10^{-3}$, 
corresponding to $a_0 (z_{\rm rec}) / f_a \lesssim 10^{-3}$ \cite{Dror:2020zru}, 
with $a_0(z_{\rm rec})$ denoting the axion field amplitude at the recombination redshift $z_{\rm rec}$. 
Expanding the axion potential around the minimum, 
$V(a) \simeq \frac{1}{2} m_a^2 a^2 - \frac{1}{4!} \lambda_a a^4 + \ldots\,$, 
the bound translates into a condition on the quartic coupling \cite{Dror:2020zru}
\beq 
\label{eq:bounddror}
\left. \frac{\lambda_a a_0^4}{m_a^2 a_0^2}\right|_{\rm eq} \sim 
\frac{\lambda_a \, \text{eV}^4}{m_a^4} \lesssim 10^{-3} \, . 
\eeq 
Ref.~\cite{Dror:2020zru} considered such constraint in the context of the 
$Z_\N$ axion model of Ref.~\cite{Hook:2018jle}. 
We here revisit the 
latter analysis with 
the formulae derived in Ref.~\cite{ZNCPpaper} for the $Z_\N$ scenario under discussion. 
For the latter, the potential in the large $\N$ limit --\cref{Eq: fourier potential large N hyper,maZNLargeN}-- corresponds to a quartic coupling given by
 \beq
 \lambda_a = m_a^2 f_a^2 / \N^2\,,
 \eeq and the bound 
in \eq{eq:bounddror}
translates into 
$\N \lesssim 85$.  This constraint does not impact the DM prospects discussed in this paper, as $\N\gtrsim65$ values  are already excluded by   either finite density constraint or nEDM data, see Fig.~\ref{fig:axionEDM}.

Finally, we mention that the pre-inflationary PQ breaking scenario considered here is subject to potential CMB constraints on iso-curvature fluctuations. Since the dynamics of the axion field evolution is highly non-linear it is not obvious to track the evolution of the iso-curvature fluctuations from the end of inflation till recombination. We note, however, that these bounds are relaxed for low-scale inflation $H_I \ll 10^{11}$ GeV (see e.g. Ref.~\cite{Alvarez:2017kar}, for an analysis based on a fine-tuned model). 

In summary, the $Z_\N$ scenario under discussion allows us to account for DM {\it and} solve the strong CP problem in a sizeable fraction of the ultra-light axions parameter space: 
$m_a \in \[10^{-22}, 10^{-10}\]$ eV.  
For instance, Fig.~\ref{fig:axionphoton} shows that a photon-axion 
signal could be at reach for the upper masses in that range, 
provided $\theta_1$ takes sizeable values and the kinetic misalignment regime takes place. 
More importantly,  
in that entire mass range 
a model-independent discovery signal is open for observation  
at the oscillating nEDM experiments such as CASPEr-Electric, 
see Fig.~\ref{fig:axionEDM}. The latter figure also points to the exciting prospects ahead, because the near-future data from high-density stellar systems and gravitational wave facilities should cover that entire mass region, for any value of  $\theta_1$ and  down to $\N\sim9$.

\section{Conclusions}
\label{sec:conclusions}

This work constitutes a proof-of-concept that an axion lighter --or even much lighter-- than the canonical QCD one may both solve the SM strong CP problem {\it and} account for the entire DM density of the Universe. 
 Large regions of the $\{m_a,\,f_a\}$ parameter space to the left of the canonical QCD axion band can accomplish that goal. 
 
 While the implications of a $Z_\N$ shift symmetry to solve the strong CP problem (with a $1/\N$ probability and $\N$ degenerate worlds) have been previously analyzed~\cite{ZNCPpaper}, leading to a lighter-than-usual axion, the question of DM and the cosmological evolution was left unexplored.
We showed here that the  evolution of the axion field through the cosmological history 
departs drastically from both the standard one and from previously considered mirror world scenarios. 

In particular, we identified a novel axion production mechanism which holds whenever 
$f_a \lesssim 3.2\times 10^{17}\, \text{GeV}$: {\bf trapped misalignment}, which is a direct consequence of  the temperature dependence of the axion potential. Two distinct stages of oscillations take place. 
   At large temperatures the minimum of the finite-temperature potential shifts from its vacuum value, i.e.~$\theta=0$,  to large values, e.g.~$\theta=\pi$, where the axion field gets trapped down to a temperature $T\sim T_{\rm QCD}$.  The axion mass is unsuppressed during this trapped period and thus of the order of the canonical  QCD axion mass. The underlying reason is that the SM thermal bath  explicitly breaks the $Z_\N$ symmetry,  because  its  temperature must be higher than that of the other mirror worlds. This trapped period has a major cosmological impact:  the subsequent onset of oscillations around the true minimum at $\theta=0$  is delayed as compared with the  standard QCD axion scenario. The result is an important enhancement of the DM relic density.
    In other words, 
    lower $f_a$ values can now account for DM.
    
    We have determined the minimum kinetic energy $K_{\text{min}}$ required at the end of trapping
     for the axion to roll over $\sim\N/2$ maxima before it starts to oscillate around the true 
     minimum (so as to solve the strong CP problem). 
    We showed that the axion kinetic energy is  of $\cal{O}( K_{\text{min}})$  in sizeable regions of the parameter space, fuelled by the (much larger than in vacuum) axion mass at the end of the trapped period.
    In this {\bf pure trapped} scenario, the final oscillations start at temperatures smaller but 
    close to $T\sim T_{\rm QCD}$.

  In fact,  the axion kinetic energy at the end of trapping  is shown to be in general much larger than $K_{\text{min}}$.
  Trapped misalignment then automatically seeds kinetic misalignment~\cite{Co:2019jts} between 
  $T\sim T_{\rm QCD}$ and lower temperatures.  
 The axion rolls for a long time over the low-temperature potential barriers before final oscillations start at $T\ll T_{\rm QCD}$, extending further the delay of oscillations around the true minimum ensured by the trapped period.   
 In consequence, the {\bf trapped+kinetic} misalignment mechanism enhances even more strongly the DM relic density.

      Our novel trapped mechanism is  more general than the $Z_\N$ framework considered here. It could arise in a large variety of ALP or QCD axion scenarios. For instance, it may apply to axion theories  in which an explicit  source of PQ breaking is active only at high temperatures and the transition to the true vacuum is non-adiabatic. 
    Note also that in our scenario kinetic misalignment does not rely on the presence 
    of non-renormalizable 
    PQ-breaking operators required in the original formulation~\cite{Co:2019jts}.
    It is instead directly seeded by trapped misalignment, which is itself a pure temperature effect.

  For values of the $Z_\N$ axion scale $f_a \gtrsim 3.2\times 10^{17}\, \text{GeV}$,  the trapped mechanism does not take place, since there is only one stage of oscillations. 
  The  $T=0$ potential is already developed
when the Hubble friction is overcome, and the axion  oscillates from the start around the true minimum $\theta_a=0$.   The   relic density corresponds then to that of a {\bf simple ALP} regime with constant axion mass, alike to the standard QCD axion scenario. 
  
  We have determined  the current axion relic density stemming from the various misalignment mechanisms, analyzing their different dependence on the $\{m_a,\,f_a\,,
   \N\}$ variables. The  ultimate dependence on  the arbitrary initial misalignment angle has been determined as well for the simple ALP and trapped+kinetic scenarios. 
   For the  pure trapped scenario, the relic density turns out to be independent of the initial misalignment, which results in a band centered around $\N\sim21$ to account for the ensemble of DM. Overall, DM solutions are found within the $Z_\N$ paradigm for any value of $3\le\N\lesssim 65$. 

 The results 
 above have been next confronted with the experimental arena of the so-called axion DM searches. As a wonderful byproduct of the lower-than-usual $f_a$ values 
 allowed in the $Z_\N$ 
 axion paradigm to solve the strong CP problem, 
  all axion-SM couplings are equally enhanced for a given $m_a$.  This increases the testability of the theory in current and future experiments. 
 In consequence, many axion DM experiments which up to now only aimed to target the nature of DM,
  are simultaneously addressing  the SM strong CP problem, provided mirror worlds exist.
   We have studied the present and projected experimental sensitivity to the axion coupling  to photons, electrons and nucleons, as a function of the axion mass and  $\N$.
    It follows that an axion-photon signal is at reach in large portions of the parameter space of upcoming axion DM experiments, while 
no such prospects result for the coupling to nucleons, and only marginally for the coupling to electrons.

  A different and crucial test  is provided by the $aG\tilde G$ coupling (that fixes the value of $1/f_a$), 
  which  can be entirely traded by an axion-nEDM coupling. The signal 
   has two remarkable properties, for any given $m_a$: i) in all generality, it {\it does not} depend on the details of the putative UV completion of the axion model, unlike all other couplings considered; ii)  
    its strength is enhanced in the $Z_\N$ paradigm, which is impossible in any model of the canonical QCD axion.
It follows that the $Z_\N$ paradigm is --to our knowledge-- the only true axion theory  that could explain a positive signal in  CASPEr-Electric phase I and in a large region of the parameter space in phase II. The reason is that a traditional QCD axion with an nEDM coupling in the range to be probed by that experiment 
would be automatically heavier, and therefore outside its reach.  Such a signal could instead account for DM  {\it and} solve the strong CP problem within the $Z_\N$ scenario. 
 The same applies to the Storage Ring projects that aim to detect  oscillating  EDMs.

Furthermore, our results demonstrate   
a beautiful synergy and complementarity between the expected reach of axion DM experiments and   axion experiments which  are independent of the nature of DM. 
 For instance, oscillating nEDM experiments on one side,  and data expected from highly dense stellar objects and gravitational waves on the other, have a wide overlap in their sensitivity reach. Their combination  will cover in the next decades the full range of possible $\N$ and $m_a$ values, in the mass range from the standard QCD axion mass down to $\sim 10^{-22}$ eV, that is, down to the fuzzy DM range. To our knowledge, the $Z_\N$ axion discussed here is the first model of fuzzy DM which also solves the strong CP problem.

\begin{small}

\section*{Acknowledgments}
We thank Gonzalo Alonso-\' Alvarez, Quentin Bonnefoy, Gary Centers, Victor Enguita, Yann Gouttenoire, Benjamin Grinstein, Lam Hui, David B.~Kaplan, D. Marsh, V. Mehta, Ryosuke Sato, Geraldine Servant, Philip S{\o}rensen, Luca Visinelli and  Neal Weiner for illuminating discussions. 
The work of L.D.L.~is supported by the Marie Sk\l{}odowska-Curie Individual Fellowship grant AXIONRUSH (GA 840791). 
L.D.L., P.Q. and A.R.  acknowledge support by  
the Deutsche Forschungsgemeinschaft under Germany's Excellence Strategy 
- EXC 2121 Quantum Universe - 390833306.
M.B.G.~acknowledges support  from the ``Spanish Agencia Estatal de Investigaci\'on'' (AEI) and the EU ``Fondo Europeo de Desarrollo Regional'' (FEDER) through the projects FPA2016-78645-P and PID2019-108892RB-I00/AEI/10.13039/501100011033.
M.B.G. and P.~Q. acknowledge support from the European Union's Horizon 2020 research and innovation programme under the Marie Sklodowska-Curie grant agreements 690575  (RISE InvisiblesPlus) and  674896 (ITN ELUSIVES), as well as from  the Spanish Research Agency (Agencia Estatal de Investigaci\'on) through the grant IFT Centro de Excelencia Severo Ochoa SEV-2016-0597. 
 This project has received funding/support from the European Union's Horizon 2020 research and innovation programme under the Marie Sklodowska-Curie grant agreement No 860881-HIDDeN.

\appendix

\section{Ratios of $g_*$ and $g_s$}
\label{App: ratios gs grho}
Throughout the paper, different ratios of the number of degrees of freedom in the SM and the mirror worlds appear. They need to be computed numerically and correspond to $\mathcal{O}(1)$ corrections that in large regions of the parameter space can be neglected. This     
allows the derivation of simple analytical formulas for the different observables, such as the contributions to $N_{\rm eff}$, or the axion relic density. In this Appendix we review the definition of these quantities and compute them numerically as a function of both the SM and mirror$_k$ temperatures for different values of $\gamma_k$. To this aim, the tabulated effective number of degrees of freedom (for the SM case) as a function of temperature provided in Ref.~\cite{Saikawa:2018rcs} will be used (see also 
\fig{Fig: dof  approximation}).
\begin{figure}[!ht]
\centering
\includegraphics[width=0.65\textwidth]{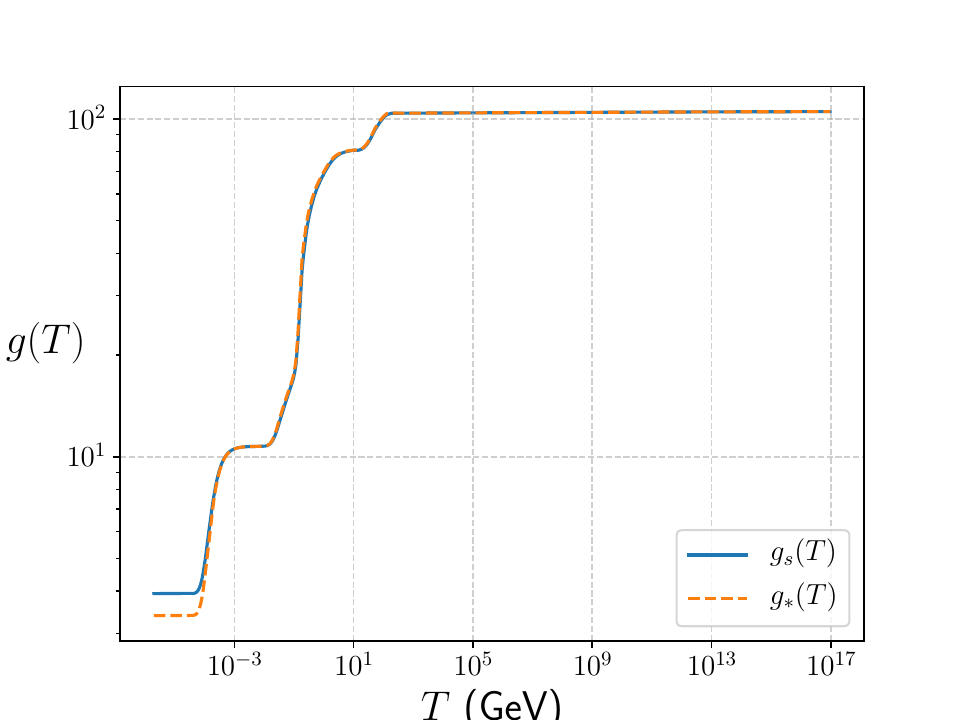}  
\caption{Effective number of degrees of freedom.}
\label{Fig: dof  approximation}       
\end{figure}
 
\subsubsection*{Ratio of temperatures vs.~$\gamma_k$: $b_k(T,T_k)$}
The function $b_k(T,T_k)$ defined in  Eq.~(\ref{gammalong}),
\begin{align}
b_k(T,T_k)\equiv \bigg[\frac{g_{s}(T)}{g_{s}\left(T_k\right)}\bigg]^{1 / 3}=\frac{T_k/T}{\gamma_k} \, ,
\end{align}
is shown to approach $1$ for $T_k\gg \LQCD$ in the right plot in \cref{Fig: dof bk}. 
Therefore the approximation $b_k(T,T_k)\sim 1$ can be safely adopted in the computation of the high-temperature axion potential in \cref{Eq: high temp fot equal T}.  
 \begin{figure}[ht]
\centering
\includegraphics[width=0.495\textwidth]{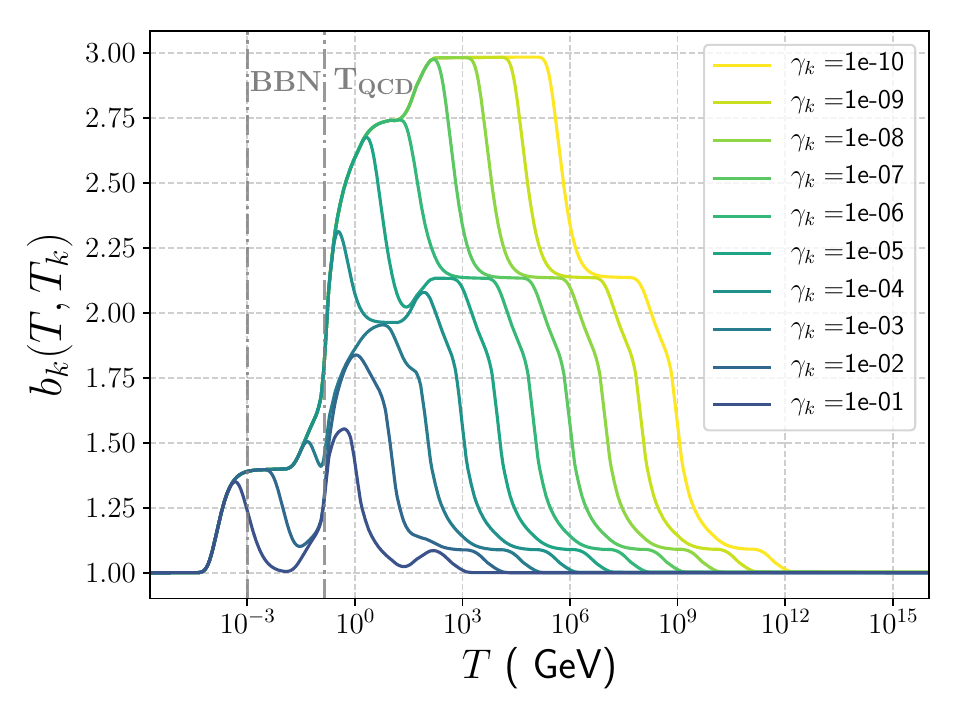}
\includegraphics[width=0.495\textwidth]{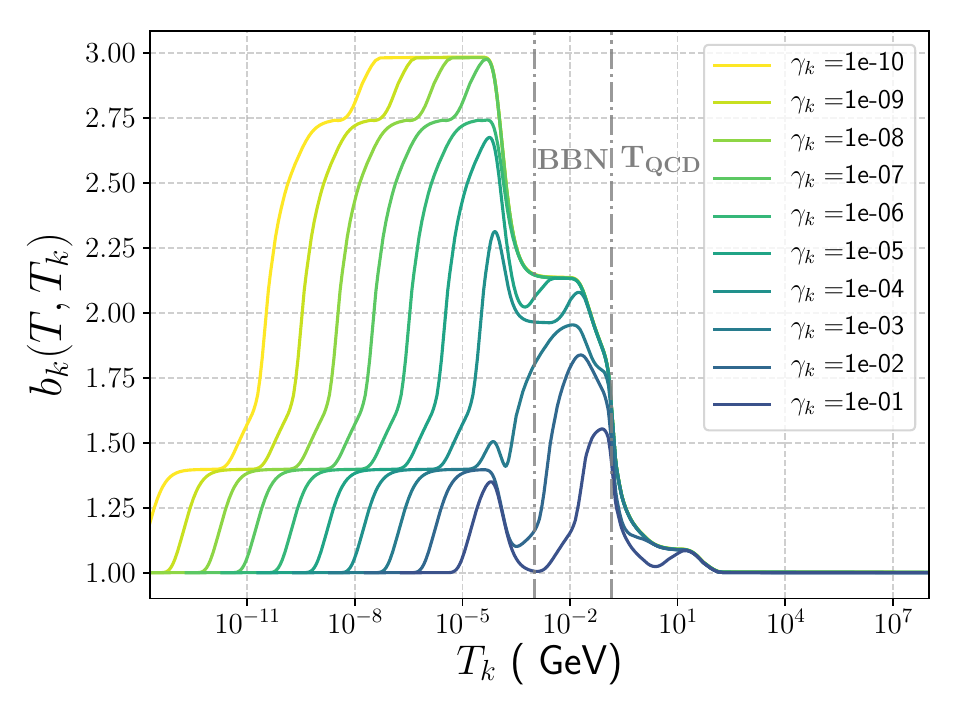}  
\caption{Numerical evaluation of  $b_k(T,T_k)$ as a function of the SM temperature $T$ (left) and of the k-th mirror world temperature $T_k$ (right), for different values of $\gamma_k$.}
\label{Fig: dof bk}       
\end{figure}

\subsubsection*{Bounds on $\gamma_k$ through $N_{\rm eff}$: $c_k(T,T_k)$}
The ratio 
\begin{align}
c_k\left(T, T_k\right)\equiv \left[\frac{g_{*}\left(T_{k}\right)}{g_{*}(T)}\right] \left[\frac{g_{s}(T)}{g_{s}\left(T_{k}\right)}\right]^{4 / 3}\,,
\end{align}
 which appears in Eq.~(\ref{Eq: Zn number of rel dof g*}), has been depicted in \cref{Fig: dof ck}. The left plot shows that its value is $c_k(T=1\text{ MeV},T_k)\lesssim 1.2$ whenever the SM BBN is taking place, i.e.~$T\sim 1\text{ MeV}$. In consequence, the approximation  $c_k(T,T_k)\sim 1$ can be safely applied when using the experimental limits in \cref{Eq: BBN neff bound} to extract bounds on $\gamma_k$  from \cref{Eq: Neff BBNMeV}. 
 \begin{figure}[h]
\centering
\includegraphics[width=0.495\textwidth]{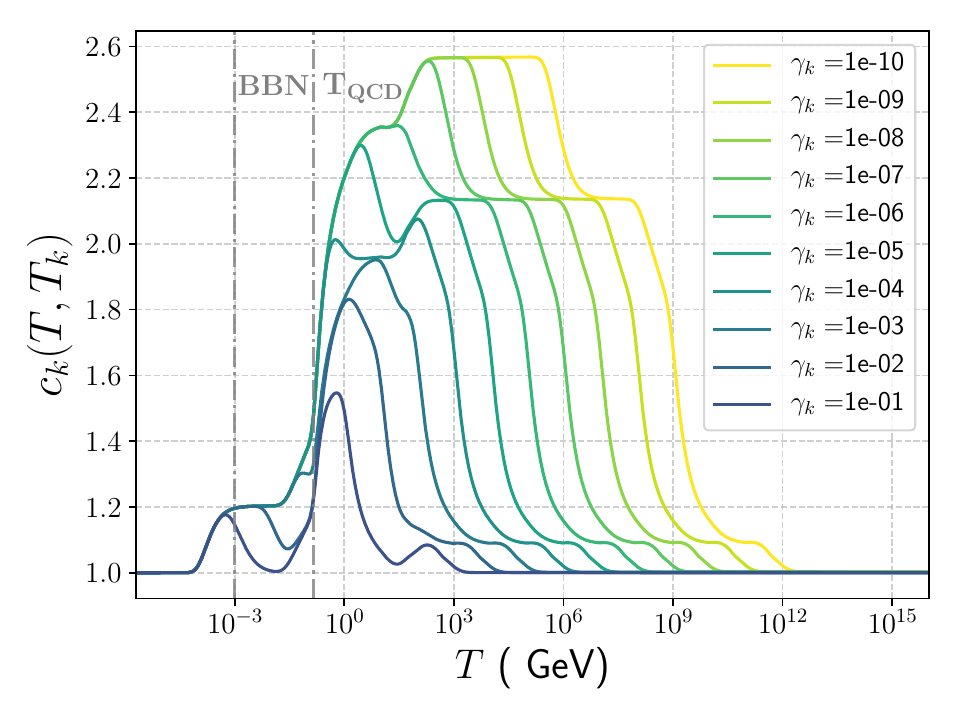}
\includegraphics[width=0.495\textwidth]{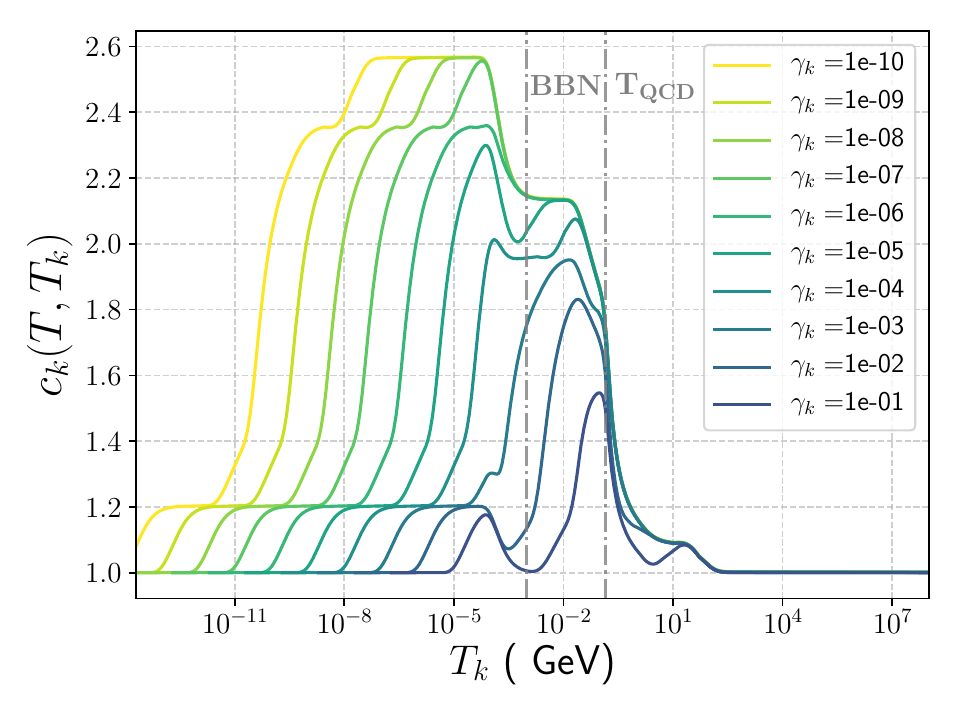}  
\caption{Numerical evaluation of  $c_k(T,T_k)$ as a function of the SM temperature $T$ (left) and  of the k-th mirror world temperature $T_k$ (right), for different values of $\gamma_k$. }
\label{Fig: dof ck}       
\end{figure}

\subsubsection*{Trapped+kinetic misalignment: $\mathcal{F}_{\rm kin,\,2}(T_{1},T_1')$}
The prediction for the relic density in the kinetic+trapped misalignment scenario in \cref{Eq: energy density Kinetic+trapped only m1-1} involves the ratio
\begin{align}
\mathcal{F}_{\rm kin,1}(T_{1})\equiv\Bigg(\frac{\sqrt{g_{*}(T_{1})g_{*}(T_{\rm QCD})}}{3.38} \Bigg)^{\frac{3}{4}}\left(\frac{3.93}{\sqrt{g_{s}(T_{1})g_{s}(T_{\rm QCD})}} \right)\,, 
\label{Eq:fkin1}
\end{align}
which feeds into the final result in \cref{Eq: energy density Kinetic+trapped equal gamma Nfa} (see also \cref{App:Maximum relic density}) through $\mathcal{F}_{\rm kin,\,2}$:
\begin{multline}
\label{Eq:fkin2}
  \mathcal{F}_{\rm kin,\,2}(T_{1},T_1')\equiv \mathcal{F}_{\rm kin,1}(T_{1})
  \times b'(T_1,T'_1)^{1/3}
  \times c'(T_{\rm BBN},T'_{\rm BBN})^{-1/12}
  \times \left(3.38/g_{*}(T_1)\right)^{1/12}=  \\
  \left[\frac{\sqrt{g_{*}(T_{1})g_{*}(\LQCD)}}{3.38}\right]^{\frac{3}{4}}
  \left[\frac{3.93}{\sqrt{g_{s}(T_{1})g_{s}(\LQCD)} }\right]
   \left[\frac{g_{s}(T_{1})}{g_{s}\left(T'_{1}\right)}\right]^{1/9}
   \left[\frac{g_{*}\left(T_{\rm BBN}\right)}{g_{*}(T'_{\rm BBN})}\right]^{1/12} \\
        \times \left[\frac{g_{s}(T'_{\rm BBN})}{g_{s}\left(T_{\rm BBN}\right)}\right]^{1/9}
    \left(\frac{3.38}{g_{*}(T_1)}\right)^{1/12}\,,
  \end{multline}
  where  $T'_{\rm BBN}$ denotes the temperature of all the mirror copies of the SM when the temperature of the latter is $T_{\rm BBN}\simeq 1\text{ MeV}$.
  
  The numerical evaluation of $\mathcal{F}_{\rm kin,\,2}$ can be found in \cref{Fig: dof Fkin2}. It shows that its  
  values lie in the range $0.45-0.85$, depending on the temperature at the onset of the first stage of oscillations, $T_1$, and on the value of $\gamma_k$.
   \begin{figure}[h]
\centering
\includegraphics[width=0.495\textwidth]{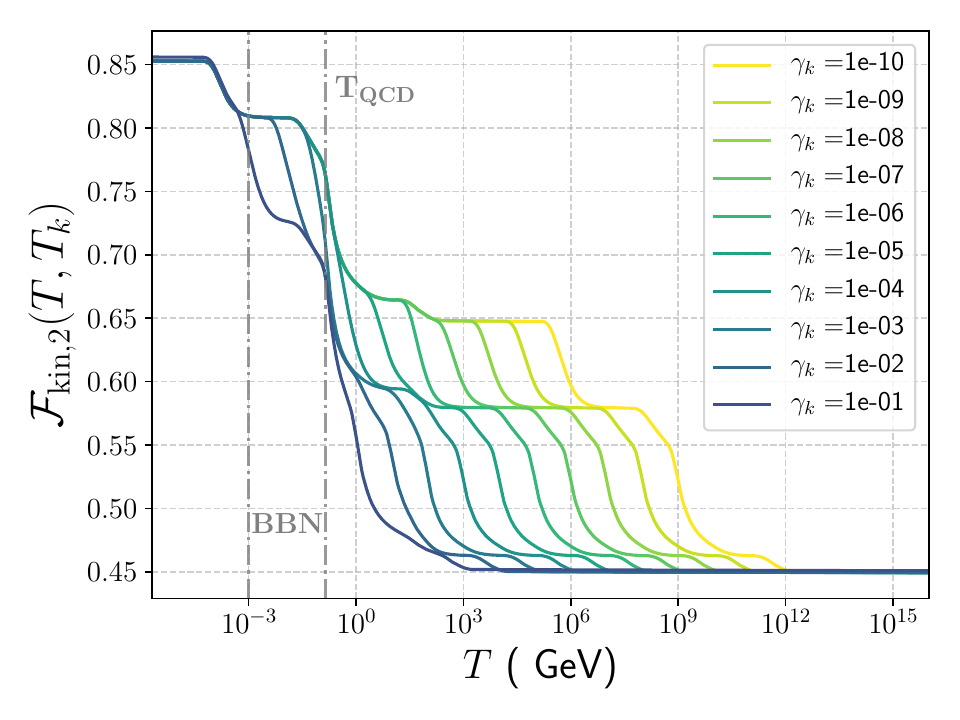}
\includegraphics[width=0.495\textwidth]{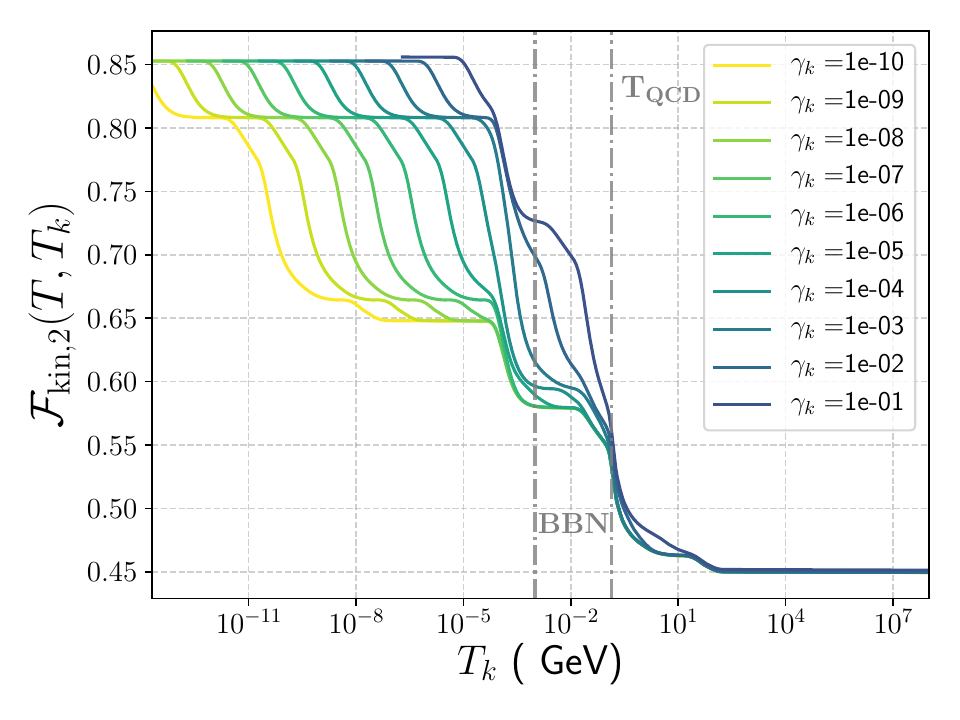}  
\caption{Numerical evaluation of $\mathcal{F}_{\rm kin,\,2}(T_{1},T_1')$ as a function of the SM temperature $T$ (left) and of the k-th mirror world temperature $T_k$ (right), for different values of $\gamma_k$. }
\label{Fig: dof Fkin2}       
\end{figure}

\section{Anharmonicity function} 
\label{sec:anharmonicity_function}
The relic density of a scalar field with a harmonic potential can be computed analytically using the adiabatic approximation. However, pseudo Goldstone bosons are in general described by 
periodic potentials that only resemble a harmonic potential close to the minimum. If the initial misalignment angle is large $\theta \gtrsim \pi/2$, the onset of oscillations gets delayed leading to an enhancement of the relic density. This enhancement can be parametrized via the so-called anharmonicity factor $f_{\rm anh}$. For an ALP-like regime, one can use   an empirical formula proposed in Ref.~\cite{Arvanitaki:2019rax} which encodes this deviation from the harmonic result:   
\begin{align}
\rho_{a,\, 0}= \frac{1}{2}m_a^2 \left(\frac{a_1}{a_{\rm 0}}\right)^3 \theta_{1}^ 2f_a^2 f_{\rm anh}(\theta_1) \, .
\end{align}
The definition of $f_{\rm anh}(\theta_1)$ thus depends  
on the criteria chosen to define the onset of oscillations at temperature $T_1$. In our case, $T_1$ corresponds to the temperature in which the axion mass term overcomes Hubble friction as given in Eq.~(\ref{T1}).  
\begin{figure}[h]
\centering
\includegraphics[width=0.7\textwidth]{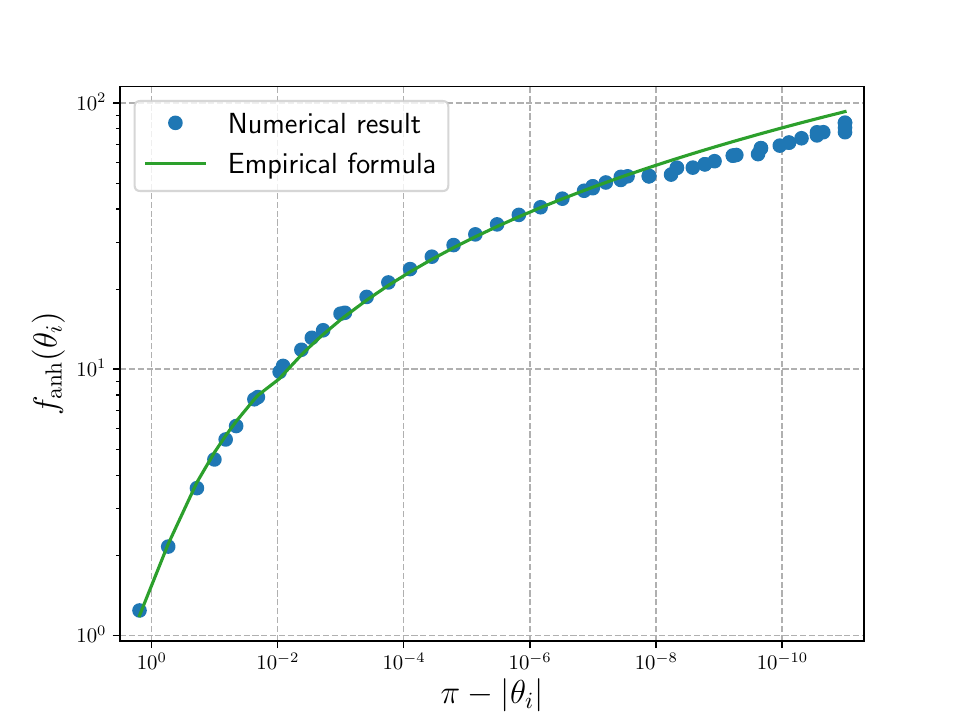} 
\caption{Comparison of the empirical formula in \eq{eq:fanhemp} for the anharmonicity factor 
with the numerical result in the simple ALP case.}
\label{fig:Anharmonicity}       
\end{figure}
For the case of a simple ALP with constant mass and a cosine potential, and for large misalignment angles $\pi-\theta_1 \lesssim 10^{-2}$, our numerical result is in excellent agreement with the Ref.~\cite{Arvanitaki:2019rax} proposal. 
However, the latter does not apply for small misalignment angles, as it does not have the expected limiting behaviour $ f_{\rm anh}\left(\theta_{1}\right)\xrightarrow{\,\,\theta_{1} \rightarrow 0\,\,}1$. For this reason, we will employ the following anharmonicity prescription: 
\begin{align}
\label{eq:fanhemp}
f_{\rm anh}(\theta_1)=\left\{\begin{array}{ll}
\left[1-\log\left({1-\left({\theta_1}/{\pi}\right)^3}\right)\right]^{1.3}
& \text{ for }\quad \pi-\theta_1 > 10^{-2}\,,\\[12pt]
0.06\left[
\log \left(\frac{2.32}{\pi-|\theta_1|}\right) +
4\,\log\left(\log \left(\frac{2.32}{\pi-|\theta_1|}\right)\right)
\right] & \text{ for }\quad \pi-\theta_1 < 10^{-2}\,,
\end{array}\right.
\end{align}
where the first expression  is inspired by  Ref.~\cite{Lyth:1991ub}, which obtained an expression for the anharmonicity factor for the QCD axion  (we have adapted it so as to fit our numerical result for the constant mass ALP); and the second expression is taken from Ref.~\cite{Arvanitaki:2019rax}.

\section{Maximal relic density in the trapped+kinetic regime}
\label{App:Maximum relic density}
The predicted relic density in the trapped+kinetic scenario of \cref{Eq:relic density a trapped+kinetic} can be rewritten as
       \begin{align}
       \frac{\rho_{a,0}}{\rho_{\rm DM}}\bigg|_{\rm tr+kin} \simeq 7.9  
       \sqrt{\frac{m_{a}}{\mathrm{eV}}} 
       \bigg(\frac{m_{a}^2\,m_{a,\pi}^{\rm QCD}}{m_{\rm crit}^3}\bigg)^{1/4} 
       \left(\frac{m_{a,\pi}^{\rm QCD}}{m_1}\right)^{1/4}
       \left(\frac{\,f_a}{ 10^{12}\, \mathrm{GeV}}\right)^{2}\,
       \frac{\left|\theta_{1}\right|}{\N}\, \mathcal{F}_{\rm kin,1}(T_{1}) \, , 
       \label{Eq: energy density Kinetic+trapped only m1}
       \end{align}
       where   $\mathcal{F}_{\rm kin,1}(T_{1})$ was given in Eq.~(\ref{Eq:fkin1}).  
       The relic density depends via $m_1$ on the full temperature-dependent potential in \cref{ZNpotentialT} and therefore on the temperatures 
       of all the different worlds (i.e.~ $T$ and the values of $\gamma_k$, see \cref{gammalong}). In the simplified case in which all the mirror copies of the SM have the same temperature $T_{k \neq 0} = T^{\prime}$ one can use the potential in \cref{Eq: high temp fot equal T} to express the factor $({m_{a,\pi}^{\rm QCD}}/{m_1})^{1/4}$ as a function of the parameters of the finite-temperature potential, 
       \begin{align}
       \left(\frac{m_{a,\pi}^{\rm QCD}}{m_1}\right)^{1/4}\simeq 
       \left(\frac{\gamma' \,b'(T_1,T_1')\,\sqrt{m_{a,\pi}^{\rm QCD} M_{\rm pl}}}{T_{\rm QCD}} \right)^{\frac{\alpha}{2\alpha +8}}
       \left(1.67\times 1.66\,\sqrt{g_{*}(T_1)}\right)^{\frac{-\alpha}{4\alpha +16}}\,.
       \end{align}
       For $\alpha=8$ it results in  a parametric dependence on the mass scales of the scenario of the form 
       \begin{align}
       \frac{\rho_{a,0}}{\rho_{\rm DM}} \propto 
       \sqrt{\frac{m_{a}}{\mathrm{eV}}} \,
       \Bigg[\frac{m_{a}^6\,\big(m_{a,\pi}^{\rm QCD}\big)^5 M_{\rm Pl}^2}{m_{\rm crit}^9 T_{\rm QCD}^4}\Bigg]^{1/12} \left(\frac{\,f_a}{ 10^{12}\, \mathrm{GeV}}\right)^{2}\,
       \frac{\left|\theta_{1}\right|}{\N}\,,
       \end{align}
       from which the parametric dependences shown in \cref{tablemafa} follow.
       
       Finally, the maximal relic density that can be generated by the trapped+kinetic misalignment mechanism can be obtained as a function of $f_a$ and $\N$, by choosing the value of $\gamma'$ that saturates the $N_{\rm eff}$ bound
       in \cref{Eq: Neff BBNMeV}.\footnote{Corresponding to $\gamma'={0.51}/{(\N-1)^{1/4}}c'^{-1/4}(T_{\rm BBN}, T'_{\rm BBN})$, with $c'\equiv c_{k\neq 0}$.
}  In the case of the $Z_\N$ axion mass in \cref{maZNLargeN} 
       we obtain
       \begin{align}
       \frac{\rho_{a,0}}{\rho_{\rm DM}}\bigg|_{\rm tr+kin,\, max} \simeq 0.64 
       \left(\frac{\,f_a}{ 10^{9}\, \mathrm{GeV}}\right)^{5/6}\,
       \frac{z^{\N/2}}{\N^{1/3}} \,\left|\theta_{1}\right|\,\mathcal{F}_{\rm kin,\,2}(T_{1}) \, , 
       \label{Eq: energy density Kinetic+trapped equal gamma Nfa App}
       \end{align}
       where   $\mathcal{F}_{\rm kin,\,2}(T_{1})$ can be found in Eq.~(\ref{Eq:fkin2}).
          
\bibliographystyle{utphys.bst}
\bibliography{bibliography}

\end{small}

\end{document}